**Indirect Influences of Artificial Intelligence Advice on Complex Sepsis Treatment Decisions**

Venkatesh Sivaraman PhD,[1] Joel Levin PhD,[2] Eric P. Mason MEng,[3,4] Victor Talisa PhD,[4] Suchi Saria PhD,[5] Jean Feng PhD,[1] Andrew J. King PhD,[6] Jeremy M. Kahn MD,[4] and Adam Perer PhD*[3]

1. Department of Epidemiology and Biostatistics, University of California, San Francisco, San Francisco, California, United States
2. Rady School of Management, University of California, San Diego, La Jolla, California, United States
3. Human-Computer Interaction Institute, Carnegie Mellon University, Pittsburgh, Pennsylvania, United States
4. Department of Critical Care Medicine, University of Pittsburgh School of Medicine, Pittsburgh, Pennsylvania, United States
5. Department of Computer Science, Johns Hopkins University, Baltimore, Maryland, United States
6. Department of Biostatistics and Epidemiology, University of Pennsylvania, Philadelphia, Pennsylvania, United States

*Corresponding author: Adam Perer; 5000 Forbes Avenue, Pittsburgh PA 15213, adamperer@cmu.edu

## Abstract

Sepsis is a major public health concern, and many sepsis patients die due to ineffective hemodynamic management. Artificial intelligence could support hemodynamic decisions, particularly in complex cases where conflicting information and comorbidities drive uncertainty and practice variation. However, these systems' real-world value depends on not just accuracy, but how they influence the quality of clinician decisions. We first retrospectively evaluated an AI system's recommendations in 6,928 sepsis patients, finding that they improved upon the original providers' decisions overall but had significantly worse performance in complex cases. We then conducted a pre-registered experiment with 228 US critical care physicians to test their use of this imperfect advice in six complex treatment decisions, using one of three advice types designed to vary reasoning autonomy. Participants given simple model recommendations overcame poor advice, choosing optimal treatments in 56.3% of decisions compared to 31.8% if they had always followed the recommendations. They also made significantly better decisions than those receiving a less prescriptive advice type (42.9%). Physicians selectively adapted and improved weak advice, likely because it prompted them to reconsider their assessment. Our findings suggest that improving sepsis care decisions with AI may require accounting for these indirect influences resulting from human-AI collaborative reasoning.

## Introduction

Sepsis is an ongoing public health concern with involvement in approximately one in every five deaths worldwide.[1] Hemodynamic management decisions, including the administration of intravenous fluids and vasopressors to support adequate end-organ perfusion, play a critical role in determining sepsis patient outcomes.[2,3] Artificial intelligence (AI) systems have the potential to improve these decisions by learning optimal treatment strategies from observational data;[4–6] however, they currently face steep barriers to implementation in practice.[7,8] Most importantly, it is unclear whether use of these systems would result in improved decisions in actual practice. Existing evaluations have either studied AI outputs in isolation or measured clinician acceptance of the advice regardless of its quality.[9–11] Given recent literature showing clinicians' often sub-optimal use of AI,[8,12,13] effective implementation requires a better understanding how AI advice influences clinicians' reasoning and the quality of the resulting decisions.

The present study addresses this question through two complementary evaluations of AI-based sepsis hemodynamic treatment recommendations, both focusing on complex cases. While previous evaluations of sepsis treatment models have assessed recommendation quality retrospectively and in aggregate,[4,5,9] recent literature suggests that clinicians would find hemodynamic decision support most valuable in cases with greater complexity. This is because in relatively straightforward cases, existing clinical guidelines are already well-established,[14] leading to limited impact of AI advice regardless of its accuracy.[11] Conversely, more complex or unusual sepsis presentations, such as patients with multiple comorbid conditions, conflicting and equivocal clinical signs, or lack of response to prior treatments, require careful reasoning to develop an individualized strategy. This can lead to greater uncertainty and delays to proper treatment.[15,16] Consequently, AI support in complex cases was explicitly favored by clinicians in our prior qualitative work[17] and has the greatest potential to improve decision quality in actual clinical practice.[18,19]

We first aimed to understand the quality of sepsis treatment recommendations in complex cases through a subgroup-level retrospective evaluation of a previously-developed model.[17] Prior literature has found that AI accuracy degrades in complex cases,[20] and the risk of unsafe sepsis treatment recommendations in unusual scenarios has been documented, but definitions of complexity are not consistent across studies.[21,22] We therefore evaluated our model's performance in four subgroups, each defined by different measures of case complexity, in a cohort of 6,928 held-out patients with probable sepsis. We then compared the estimated value associated with accepting our model's recommendations against that of the original prescribing clinicians' actions in each subgroup. This evaluation found that, consistent with prior work, the model's recommendations were associated with better outcomes compared to the clinicians' real-life actions. However, we also saw significant reductions in performance for patients for whom the set of viable treatments varied widely, and for patients with many abnormal clinical signs. This implied that in cases where clinicians might most need support, the quality of the recommendations appeared lower and the potential benefit of AI advice might be less certain.

Given this tension, we next sought to understand whether clinicians could still benefit from AI advice in complex cases even if it was imperfect. We conducted a pre-registered decision-making experiment in which 228 critical care physicians made hemodynamic management decisions on six complex case vignettes. We manipulated advice format to test different levels of clinician autonomy in reasoning, given recent findings that less prescriptive forms of advice could mitigate over-reliance by forcing physicians to make an independent choice.[23,24] Participants in the study were thus randomized to receive either a direct Recommended Treatment, multiple Options to Consider, or a set of Options to Avoid. The primary outcome of the study was the quality of physicians' decisions, measured by consensus ratings of a panel of sepsis experts. We also evaluated both direct (e.g., full acceptance of advice) and less direct AI influences on decisions as observed in recent qualitative work,[11,17,25] by comparing physicians' choices to AI-mentioned treatment plans on a 2-dimensional treatment grid.

Consistent with the retrospective evaluation, the AI system tended to perform worse than clinicians in the complex case vignettes we tested (although physicians also made sub-optimal decisions a majority of the time). None of the AI conditions led to significant differences in decision quality compared to unaided clinicians. However, physicians given a single Recommended Treatment made significantly better decisions than those shown Options to Avoid, and made far better decisions than would be expected if they had always followed the recommendation. Secondary analyses showed that rather than simply ignoring bad advice, participants overcame weaker recommendations by selectively adjusting them into higher-quality plans. This finding supports the existence of an indirect influence on decisions we term “advice adaptation,” which may result from the ways clinicians use AI outputs in their reasoning. Our results suggest that the true clinical impact of sepsis treatment models and other AI decision support systems may not be fully captured by retrospective evaluations of their output quality, and that designing for and evaluating these indirect influences could improve the effectiveness and clinician adoption of AI-based decision support.

## Results

This study presents two evaluations of clinical decision quality in complex cases with an existing AI system: a retrospective subgroup-level analysis, followed by a prospective, pre-registered, randomized evaluation of clinician decision-making performance using the AI advice.

### Retrospective evaluation reveals AI underperformance in complex cases

The AI system used in this study was previously trained on a subset of de-identified patient data from the MIMIC-IV dataset,[26] comprising 21,065 patients with probable sepsis, identified using standard consensus definitions. It was designed to combine the quantitative rigor and interpretability of simple prediction models with the flexibility of modern machine learning architectures. First, a deep learning model encoded each patient's demographics, comorbidities, charted observations, medications, and prior hemodynamic management decisions, aggregated at four-hour epochs. By comparing each patient to its nearest neighbors according to the model's learned representation (see Figure 1(a)), the model calculated in-hospital mortality estimates for twelve possible treatment plans on a two-dimensional choice grid spanning volume (IV fluids and diuretics) and vasopressors. The model then ranked treatments for the next epoch by their mortality risk, and recommended the treatment with the lowest risk.

We estimated the quality of this model's recommendations on a cohort of 6,928 patients in MIMIC-IV who were not seen during training, using a widely-used evaluation method in treatment recommendation studies called weighted importance sampling (WIS).[4,5,9] This method calculates an expected value for a given treatment policy, where the value represents a combination of both final discharge outcomes and epoch-level changes in illness severity. A positive difference in policy value between the AI and the original prescribing clinician indicates that the AI treatment policy is associated with better outcomes. As shown in Figure 1(b), the overall value of following the AI's treatment policy was significantly higher than that of the original prescribing clinician (value difference between AI and clinician policy, 2.97, 95% CI, 1.22 to 4.61, p = 0.003). Furthermore, the recommendations were often consistent with clinical practice, matching the original clinicians' volume and vasopressor decisions a majority of the time (67.6% for volume decisions, 90.5% for vasopressor decisions).

However, we identified significant reductions in the benefit of agreeing with the AI advice when the cohort was broken down into subgroups by four measures of case complexity. For example, in complex cases defined by a high degree of variation in the viable treatments, the AI recommended treatment policy had directionally lower value than that of the original prescribing clinician (policy value difference, -1.43, 95% CI, -4.21 to 1.83, p = 0.72), representing a significant decrease compared to all other cases (difference in differences, -4.83, 95% CI, -8.11 to -1.29, p = 0.026). The results were similar in complex cases defined by abnormal clinical values in several organ systems (policy value difference, -2.98, 95% CI -6.24 to 1.38, p = 0.43; difference in differences, -6.00, 95% CI, -9.73 to -1.42, p = 0.026). When there was a large number of viable treatment choices, the AI policy value was again directionally lower than the clinician policy value (-1.70, 95% CI, -4.97 to 3.72, p = 0.72), though this was not a significant reduction in performance (difference in differences, -4.78, 95% CI, -8.56 to 0.82, p = 0.09). Notably, for patients with high illness severity measured by the Sequential Organ Failure Assessment (SOFA) score, the policy value difference was close to that of the overall population (policy value difference, 2.37, 95% CI, -1.28 to 5.37, p = 0.49), with no significant difference between more severe and less severe cases (difference in differences, -0.14, 95% CI, -4.26 to 3.19, p = 0.94). This latter finding may suggest that the model's underperformance was due to actual complexity rather than simply greater illness severity.These results showed that despite achieving strong results using a commonly-used retrospective evaluation strategy, this model could still produce low-value recommendations under challenging conditions when they might be most needed.

## Physicians often perform sub-optimally when managing complex cases

Given the insight that our model's outputs were high-quality overall but underperformant in complex cases, we designed a randomized experiment to evaluate how this imperfect AI advice would influence physicians' decision-making. We also aimed to understand if presenting the AI advice in ways that preserve clinician autonomy in reasoning would limit over-reliance on potentially bad advice. Therefore, following previous literature on AI designs that promote physician adoption and reasoning,[23–25] we defined three advice types: a single Recommended Treatment, a set of Options to Consider (the 2-3 treatments with the lowest predicted mortality risk) and a set of Options to Avoid (the 1-3 treatments with the highest mortality risk). Intuitively, these types of advice would support high-quality decision making in the setting of imperfect advice by increasing the chances that at least some presented information is accurate, and giving physicians more latitude to incorporate their own assessments.

A total of 247 physicians participated in the study from 15 October 2025 to 4 December 2025, from whom 228 complete responses were included in the analysis. Participants were randomized to receive one of the three advice types. Each participant made a hemodynamic management decision for between four and six complex case vignettes, with the AI advice shown for all but one randomly selected case (allowing us to measure the performance of unaided clinicians). All possible treatment decisions for each case were assigned a consensus quality score prior to study recruitment by a panel of five sepsis experts, enabling us to empirically rate the quality of the participant's decisions against a reference standard.

As shown in Table 1, most participants had a base specialty of pulmonary internal medicine, an academic affiliation, and a high degree of clinical experience. The median study completion time was 18 minutes (IQR, 13 - 28 minutes), with a median time to decision of 106 seconds per case (IQR, 73 - 159 seconds).

We found that clinicians' performance without AI advice was highly variable and often sub-optimal, confirming the highly complex nature of the cases. Unaided clinicians selected optimal treatment choices only 47.1% of the time (106 out of 225 decisions; 95% CI, 42.1% to 53.1%). They selected an average of 7.7 (out of 14–16) distinct treatment choices for each case, indicating high practice variation.

### More prescriptive AI advice counter-intuitively yields higher human-AI decision quality

We found that the AI advice type had a significant effect on decision quality (p = 0.002). As shown in Figure 2, physicians receiving the Recommended Treatment had significantly higher decision quality than those receiving Options to Avoid (optimal treatment rate 56.3% with the Recommended Treatment compared to 42.9% with Options to Avoid; adjusted odds ratio [OR] 2.08; 95% CI, 1.39 to 3.12; p = 0.002). Physicians receiving the Recommended Treatment were directionally but non-significantly more likely to choose optimal treatments than those who saw Options to Consider (optimal treatment rate 47.9%; adjusted OR, 1.56; 95% CI, 1.04 to 2.34; p = 0.14).

None of the three AI information types resulted in significantly different decision quality compared to unaided clinicians; however, decision quality with the Recommended Treatment was directionally higher than that of unaided clinicians (adjusted OR, 1.69; 95% CI, 1.10 to 2.58; p = 0.077). Performance was very similar to that of unaided clinicians in the Options to Consider group (adjusted OR, 1.08; 95% CI, 0.72 to 1.63; p = 0.98) and the Options to Avoid group (adjusted OR, 0.81; 95% CI, 0.54 to 1.22; p = 0.74).

Notably, the expected rate of optimal decisions would have been markedly different than observed had clinicians always followed the advice they received, as shown by the gray bars in Figure 2. Always following Recommended Treatment and Options to Consider would have resulted in the lowest-quality decisions, with 31.8% (95% CI, 28.8% to 34.5%) and 31.2% (95% CI, 27.8% to 34.0%) of decisions reaching optimal quality, respectively. Always following Options to Avoid would have yielded decision quality similar to unaided clinicians at 43.2% (95% CI, 38.1% to 47.8%), reflecting its less prescriptive nature. Given the low expected performance of always following Recommended Treatment but its relatively high observed decision quality, participants in this condition may have been able to overcome poor advice more effectively than those who received other types of AI information.

### Choosing adjacent treatments under poor advice helps preserve decision quality

We next conducted a secondary analysis to understand how participants' alignment with AI advice changed when they saw the advice in each condition, and whether alignment was selective based on advice quality. Using the two-dimensional treatment choice grid of volume and vasopressors (example shown in Figure 3(d)), decisions were categorized as being "Mentioned" by the AI, "Adjacent" to a mentioned plan (i.e., at most one space away), or neither of the two, categorized as "Other." We analyzed each type of alignment as a two-way interaction between whether the decision was made with AI or not, and whether the advice was good- or poor-quality for the given case and AI condition (defined by its expected effect on optimal treatment rates). This analysis showed strikingly different patterns for each advice type, depicted in Figure 3(a)-(c).

Participants who saw the Recommended Treatment displayed tendencies toward both relying on poor advice and shifting towards Adjacent plans. They chose plans Mentioned by the AI 18.8 percentage points more often than unaided clinicians (8.0% to 26.8%, $p < 0.001$). The degree of shift towards Mentioned plans was not significantly different when the advice was good compared to when it was poor, indicating potential unwarranted reliance on the AI advice. Meanwhile, the rate of Adjacent plans was similar to that of unaided clinicians overall (33.3% to 33.2%, $p > 0.99$). However, participants chose Adjacent plans significantly less often than unaided clinicians when AI advice was good (48.7% to 21.3%, $p = 0.003$), and directionally *more* often when the advice was poor (25.5% to 38.8%, $p = 0.09$). These shifts were significantly different in magnitude (adjusted difference-in-differences OR = 6.64; 95% CI, 2.90 to 15.2; $p < 0.001$), indicating that participants chose Adjacent treatments selectively based on the quality of the advice.

Aligning with the AI partially also had benefits on decision quality, helping explain the higher rate of optimal treatments in the Recommended Treatment group. As shown in the lower panel of Figure 3(a), participants' decisions were optimal significantly more often when they adopted Adjacent plans than Other plans when advice quality was poor (adjusted OR = 48.4; 95% CI, 13.6 to 172.6; $p < 0.001$).

The patterns of alignment in the Options to Consider and Options to Avoid conditions were markedly different and more consistent with over-reliance (Figure 3(b) and (c); full details in Supplementary Section S.4.2). Participants who saw Options to Consider chose Mentioned plans significantly more often (15.6% to 38.5%, $p < 0.001$), and Adjacent plans directionally less often (47.1% to 36.1%, $p = 0.058$). There were no significant differences in alignment between poor- vs good-quality advice, indicating that participants were not as able to discern advice quality and defaulted to relying on Mentioned plans. Meanwhile, participants who saw Options to Avoid had limited shifts in alignment except for a significant decrease in Mentioned plans when its advice was poor, indicating participants relied on AI advice even when it suggested avoiding a high-quality treatment (15.5% to 4.1%, $p = 0.015$).

**Physicians report being influenced to assess cases more carefully with discordant recommendations**

Finally, we conducted a qualitative analysis to understand how participants perceived the AI's role in their decision-making (353 total free-text responses by 158 respondents). Seven participants wrote that the AI advice encouraged them to think harder about or re-evaluate the patient case, the majority of whom (5 of 7) were in the Recommended Treatment condition. One participant noted, "[The AI] made me spend more time thinking about the patient, since I felt like someone else was watching over me judging my decision." This theme was corroborated by participants' self-reported influences of the AI after each case, analyzed in Supplementary Section S.4.4.

In contrast, a theme mostly observed in the Options to Consider group (expressed by 6 participants, 5 within this condition) was confusion and discomfort when the AI suggested multiple divergent alternatives: "It is frustrating to get opposing recommendations with similar confidence." Finally, in the Options to Avoid group, many participants expressed that they would prefer to see what to do, rather than what *not* to do (23 participants).

These themes suggest that the two Options interfaces induced greater cognitive burden and displaced attention away from the patient case; as one participant put it, providing more detailed non-prescriptive information "may decrease efficiency in ordering to decipher why it made the decisions it did." On the other hand, the simplicity of the Recommended Treatment may have allowed participants to spend their cognitive effort on reasoning about the patient case itself.

## Discussion

This study evaluated the effects of AI advice on clinicians' hemodynamic management decisions in particularly complex sepsis cases. Previous studies in sepsis treatment have evaluated AI recommendation quality in aggregate and without clinician involvement, not considering how human use of the outputs might affect the model's impact on patient outcomes.[4,5,9] This study therefore advances our understanding of AI-assisted decision-making in sepsis in two ways: first, we show that both clinicians and AI recommendations are often sub-optimal in complex cases, creating a need for effective human-AI collaboration. Second, we characterize how human-AI collaborative reasoning can result in clinicians adapting advice rather than accepting or ignoring it, helping overcome AI underperformance under certain advice formats.

Complex sepsis cases require clinicians to reason about competing diagnoses, ambiguous responses to prior interventions, and multiple plausible paths forward. Compared to typical cases, where AI advice may largely serve as a proxy for existing guidelines, complex cases are a more promising area for AI support. However, despite our model's strong performance overall, its outputs in complex cases were less reliable than in other cases. This underperformance was most pronounced in cases where the patient had many abnormal clinical signs and where the model identified multiple plausible but divergent treatment options, both of which likely made the optimal treatment harder to predict. Models with better performance and larger training datasets could still fail to improve complex case performance due to the inherent uncertainty in estimating mortality risk in these subgroups using readily available clinical signals.[26] Therefore, integrating AI outputs with case-by-case human clinical judgment may be the best strategy to improve outcomes in complex cases.

Prior studies of AI-assisted clinical decision-making have focused on the risks of over-reliance on poor advice[13] or under-reliance on accurate advice,[8] assuming that clinicians would either accept or ignore recommendations. However, our results suggest that clinicians integrated the AI advice as another signal in their reasoning, which led to *indirect* influences on decisions when they adapted rather than fully accepting recommendations. Our secondary analyses showed that advice adaptation likely occurred when clinicians reconsidered their assessment in the face of discordant advice, then shifted their plan towards the advice without accepting it completely. For example, clinicians tended to accept volume plans while adjusting the vasopressor plan (see Supplementary Section S.2), a pattern consistent with prior work[11] that reflects greater equipoise in fluid administration. These findings suggest future AI implementation studies should move beyond overly simple “advice followed/advice not followed” paradigms and measure the influences of AI systems on reasoning, which can result in partial acceptance or other measurable shifts in decisions. These studies should also directly test cases in which both clinicians and AI are often sub-optimal, to ensure that AI advice does not adversely affect decisions in such scenarios.

Our results also provide evidence that retrospective *in silico* evaluation alone is insufficient to assess the viability of AI-based clinical decision support. Estimates of AI output quality in isolation effectively simulate settings in which the AI makes decisions autonomously or clinicians uncritically follow the advice. However, clinicians’ use of the advice can shift decisions in both positive and negative directions, leading to large performance differences between AI-only and human-AI performance.[8] Our study corroborates this: clinicians who saw the Recommended Treatment made better decisions than would have been expected if they always followed its advice, while in other conditions imperfect advice had no benefit. As modern AI systems grow in complexity and capability, the way they present advice becomes even more critical to avoid over-reliance and de-skilling.[27] Therefore, in line with recent calls in the healthcare AI literature,[28,29] it is imperative that researchers evaluate their systems’ effects on human decisions rather than simply measuring advice quality in isolation.

In our study, advice adaptation was most common and beneficial in the most direct AI condition, a single Recommended Treatment. Less prescriptive advice may have required more cognitive effort to comprehend, leading to more limited effects on reasoning. This finding goes against the expectation based on previous literature that more indirect advice would better preserve physicians' reasoning autonomy,[17,23] suggesting that the driving factor in whether imperfect advice can support reasoning is the simplicity of its presentation. Subsequent AI designs could adopt a "progressive disclosure" approach, showing the most direct information first and allowing the user to explore more options if desired. They could also visually flag when cases are identified as complex using the measures defined in this study, and present advice differently in these cases to encourage productive influences on reasoning.

One limitation of this experiment is that we chose to focus the decision-making experiment on complex cases rather than a broader sample of sepsis patients. This choice was intentional, since the effect of AI recommendations in simple cases would likely be modest or even negligible except with extremely large sample sizes. That said, it remains unclear how the advice types we studied would have impacted human-AI decision quality in more straightforward cases. It is also important to note that both evaluation metrics used in this study, i.e., retrospective mortality estimates and expert panel consensus ratings, are imperfect proxies for the true outcome of interest (patient survival following the treatment plan). Retrospectively estimating the quality of sepsis treatment decisions is challenging, particularly given the difficulty of attributing outcomes to individual decisions over a long ICU course.[30] Expert panel scores reflect the best clinical knowledge available and have been widely used in human-AI clinical decision-making research,[31,32] motivating their use in our study. Finally, our study recruited US-based physicians who were predominantly academic, based in pulmonary internal medicine, and willing to participate in a study related to AI. Further research is needed to examine whether physicians in other specialties, regions, and practice settings exhibit similar patterns of AI influence on clinical decisions.

## Methods

### Data and model preparation

The AI advice in this study was generated using a previously-developed transformer encoder model.[17] The training data for the model was derived from the MIMIC-IV dataset[33] and comprised 21,065 patients with possible sepsis, with an average of 30 four-hour epochs per patient. The inputs to the model included demographics, comorbidities, charted observations, medications, and prior hemodynamic management decisions for each patient-epoch. The model was trained on an unsupervised "de-noising" task, meaning it learned to reproduce observed patient data given corrupted versions of the same data (without any specific target label).

After training, the model was used to generate vector representations of all patients in the dataset, allowing us to retrieve data on patient-epochs with similar representations. We split the 100 nearest neighbor cases for each patient-epoch by the treatment they received. Treatments throughout this study were represented on a two-dimensional choice grid spanning management of intravascular volume and management of vasopressors. The model supported four possible volume decisions (none, up to 1 L of IV fluid, more than 1 L of IV fluid, or diuretics) and three vasopressor decisions (none, one vasopressor, or multiple vasopressors). We estimated mortality risk for each treatment as the empirical rate of in-hospital mortality among the neighbors for whom the treatment was given. Then, the model's recommendation was simply the treatment with the lowest estimated mortality risk.

### Retrospective evaluation

We assessed the value of following this model's recommended treatment policies using a technique called Weighted Importance Sampling (WIS), which is widely used in treatment optimization and reinforcement learning.[4,5,9] WIS enables evaluation of a policy (the AI recommendations) on a dataset of trajectories collected by a different policy (the original prescribing clinicians' treatments). Conceptually, WIS works by upweighting epochs where the true decision and the AI recommendation were consistent and downweighting others. The value of the policy is based on a "reward" function, which we defined exactly as in prior work[6] by combining large terminal rewards dependent on the mortality outcome and smaller intermediate rewards based on changes in the patient's Sequential Organ Failure Assessment (SOFA) score. We computed the difference in policy value between the AI's and the clinicians' policies on a held-out cohort of 6,928 patients from MIMIC-IV, using 10,000 bootstrapped samples to compute the confidence interval and p-value. (See Supplementary Section S.1.2 for further details.)

Then, we defined four interrelated criteria to identify complex cases, each measuring different components of complexity as perceived by physicians. The first two criteria were external to our model: high SOFA score (defined as SOFA of 10 or higher) and abnormal clinical values (defined as having a clinical value outside the normal range in at least five of the following systems: cardiac, coagulation, arterial blood gases, complete blood count, liver function, chemistries, kidney function, and neurological function). The latter two criteria were based on whether our model identified a large number of plausible treatment options for the decision (at least four candidate treatment plans) or highly-varying options (a spread of at least two spaces on the treatment grid). These model-based criteria aligned with the intuitive clinical idea of multiple disparate treatment approaches being plausible due to conflicting patient factors. We assessed the policy value difference in subgroups defined by whether each criterion was met in at least 20% of epochs. We computed confidence intervals and p-values within each subgroup using the same method as above, and we computed a difference-in-differences confidence interval and p-value between more complex and less complex cases. We corrected for multiple hypothesis testing across the four criteria using the Holm method.

## Decision-making study design

We next designed and conducted a randomized controlled vignette-based experiment to test decision quality when physicians reviewed our AI outputs in complex cases. In the study, critical care physicians made AI-assisted hemodynamic management decisions on 4–6 patient cases selected according to the above complexity criteria. For each case, participants selected a treatment plan to apply over the next four hours, choosing between 14 and 16 options on the two-dimensional choice grid. The primary outcome of the study was decision quality, measured by consensus ratings from a panel of sepsis experts.

Participants were randomized to one of three distinct AI advice types developed using the nearest neighbors and mortality risk estimates from our model: “Recommended Treatment”, “Options to Consider” and “Options to Avoid” (Figure 5). Recommended Treatment displayed the single treatment action for which similar patients receiving the treatment had the lowest mortality risk. Options to Consider displayed 2-3 treatments with lower mortality risk than other viable options, similar to prior work.[17,23] Finally, Options to Avoid showed treatment options that had the highest mortality risk, under the hypothesis that clinicians would second-guess themselves and consider the patient more carefully if they were considering a plan labeled high-risk.

Participants were recruited from two mutually exclusive pools of critical care physicians: a list of 594 US-based physicians derived from the American Medical Association Masterfile and used in prior studies,[34–36] and a set of public email addresses for 1,097 academic physicians from 30 US institutions. Participants were offered 75 USD as compensation, and the study was approved by the University of Pittsburgh institutional review board (#22040045).

This study was pre-registered at https://osf.io/h5urm and follows the Consolidated Standards of Reporting Trials (CONSORT) reporting guideline. Protocol details are provided in the supplementary materials, and the participant flowchart is shown in Figure 4.

### Case selection

Six complex case vignettes were selected for the study from real patients in MIMIC-IV. We used multiple criteria to identify complex cases, including those used in the retrospective evaluation as well as manual searches for “classic” complex sepsis patients (e.g., “patients with borderline blood pressure and a heart failure history who have recently received a lot of IV fluid”). We wrote detailed clinical vignettes for each selected case (see Supplementary Section S.5), incorporating structured data used by the model as well as context from the patients’ discharge notes such as physical exams and imaging. Each case was associated with 14–16 treatment choices, including the twelve used by the model and 2–4 conditional treatments (e.g., “Give less than 1 L of fluid, and start a vasopressor if there is no response to the fluid”).

### Outcomes

The primary outcome was decision quality, determined by consensus scores from a panel of five practicing critical-care physicians. Panel members had 15–22 years’ ICU experience and were board-certified in critical care medicine by the American Board of Internal Medicine. Quality ratings were provided for each possible response for each case on a 7-point Likert scale where 1 meant the decision would “most likely harm the patient” and 7 meant the decision would “most likely help the patient.” Panel members were not shown any AI outputs during the rating process. We considered the treatments that were assigned the maximum quality scores to represent the best possible decisions that could be made with the information available and with sufficient prior clinical knowledge and deliberation. We therefore defined decision quality throughout the analysis as the rate at which participants selected one of these “optimal” choices. Between one and four treatments were rated as optimal per patient and at least three were rated as somewhat likely to help the patient, indicating high clinical equipoise.

Secondary outcomes included participants' level of alignment with the AI advice and their perceived influences of the AI on their decisions. Alignment with AI advice was categorized as "Mentioned," "Adjacent," or "Other" based on proximity to the plans listed by each AI system. Adjacent plans were defined as being identical to a mentioned plan along one treatment dimension and consecutive along the other (see Figure S2). Participants reported the influence of the AI on their decision-making by selecting one or more of eight influence patterns derived from prior work.[17]

**Procedures**

The experiment was implemented on a commercially available online study platform (Qualtrics). Randomization was performed using the study platform, and participants were not informed about the other advice types. Participants reported their demographic information at the start of the study. During the decision-making phase, they reviewed four vignettes chosen at random. They received advice from the AI they were assigned to on all but one randomly selected case, allowing us to observe the decision-making behavior of unaided clinicians. Upon completing four cases, they were given the option to make decisions on the final two vignettes. Finally, participants were asked open-ended questions to elicit their perceptions of the AI.

**Statistical analysis**

The target sample size of 250 participants was pre-specified using a power analysis based on effect sizes in our prior work.[17] Using a proportional-odds model on the decision quality scores, the power analysis showed more than 80% power to detect a difference in at least one treatment arm with an odds ratio of 1.8 or higher, with intraclass correlations ranging from 0.1 to 0.4 and a two-sided $\alpha$ value of 0.05.

Our analysis included responses for all participants who completed the online study. The primary outcome of the analysis was the rate at which participants selected optimal treatments. (Our pre-registered analysis directly used the quality score ratings, which are more complex to interpret given their ordinal nature. The results for this analysis are shown in Supplementary Section S.4.1, with consistent hypothesis testing results to the findings shown in the main text.) To assess differences in decision quality, we fit mixed-effects logistic regression models with AI condition as a fixed effect; we included random effects to account for clustering by patient case and respondent. We tested whether the coefficients of the three AI conditions were different from each other using a Wald test, followed by Tukey honestly significant difference (HSD) testing for pairwise differences between the AI conditions and unaided clinicians. In an exploratory analysis, we compared the observed decision quality to simulated quality scores that would have resulted from always following each AI system's advice, computed by sampling decisions that agreed with the AI from unaided clinicians' responses.

To analyze alignment with AI advice, separate mixed-effects logistic regression models were fit for each alignment type. We also split the patient cases into those with "good quality" or "poor quality" AI advice based on the simulated decision quality scores above. The presence of AI and advice quality were included as fixed effects in each model (p-values corrected with the Holm method). Differences in optimal treatment rate between Adjacent and Other decisions were analyzed using logistic regression. All statistical analyses were conducted using R. Finally, open-ended responses were qualitatively analyzed by two researchers using inductive thematic analysis.

**Data availability**

The MIMIC-IV dataset is publicly available to credentialed users at https://physionet.org/content/mimiciv/3.1/. Code used to develop the model and perform the retrospective evaluation is available at https://github.com/cmudig/sepsis-reasoning-dashboard. Transformed data and model weights will be provided by the corresponding author upon reasonable request and proof of MIMIC-IV access credentials. Full deidentified survey responses and analysis code can be found at https://github.com/cmudig/sepsis-survey-analysis.

## References

1. Rudd KE, Johnson SC, Agesa KM, Shackelford KA, Tsoi D, Kievlan DR, et al. Global, regional, and national sepsis incidence and mortality, 1990–2017: analysis for the Global Burden of Disease Study. The Lancet. 2020 Jan 18;395(10219):200–11. doi:10.1016/S0140-6736(19)32989-7.
2. Jia L, Wang P, Li C, Xie J. The efficacy and safety of vasopressors for septic shock patients: A systematic review and meta-analysis. Shock. 2023 Aug 4;60(6):746–52. doi:10.1097/SHK.0000000000002193
3. Zampieri FG, Bagshaw SM, Semler MW. Fluid Therapy for Critically Ill Adults With Sepsis: A Review. JAMA. 2023 Jun 13;329(22):1967–80. doi:10.1001/jama.2023.7560
4. Kalimouttou A, Kennedy JN, Feng J, et al. Optimal Vasopressin Initiation in Septic Shock: The OVISS Reinforcement Learning Study. JAMA 2025;333:1688–98.
5. Komorowski M, Celi LA, Badawi O, Gordon AC, and Faisal AA. The Artificial Intelligence Clinician learns optimal treatment strategies for sepsis in intensive care. Nature Medicine 2018;24:1716–20.
6. Nanayakkara T, Clermont G, Langmead CJ, and Swigon D. Unifying cardiovascular modelling with deep reinforcement learning for uncertainty aware control of sepsis treatment. PLOS Digital Health 2022;1:e0000012.
7. Budzyń K, Romańczyk M, Kitala D, Kołodziej P, Bugajski M, Adami HO, et al. Endoscopist deskilling risk after exposure to artificial intelligence in colonoscopy: a multicentre, observational study. The Lancet Gastroenterology & Hepatology. 2025 Oct;10(10):896–903. doi:10.1016/S2468-1253(25)00133-5
8. Goh E, Gallo R, Hom J, et al. Large Language Model Influence on Diagnostic Reasoning: A Randomized Clinical Trial. JAMA Network Open 2024;7:e2440969–e2440969.
9. Tang Z, Sun P, Liang R, Tong S, Yang Y, Yang M, et al. Reinforcement learning for treatment decision-making in sepsis: a scoping review. npj Digit Med. 2026 Jul 18. doi:10.1038/s41746-026-03034-1

10. Nagendran M, Festor P, Komorowski M, Gordon AC, Faisal AA. Quantifying the impact of AI recommendations with explanations on prescription decision making. npj Digit Med. 2023 Nov 7;6(1):206. doi:10.1038/s41746-023-00955-z
11. Sivaraman V, Bukowski LA, Levin J, Kahn JM, Perer A. Ignore, Trust, or Negotiate: Understanding Clinician Acceptance of AI-Based Treatment Recommendations in Health Care. In: Conference on Human Factors in Computing Systems - Proceedings. Association for Computing Machinery; 2023. doi:10.1145/3544548.3581075
12. Ayyub QI, Ayesha A, Ullah KA, Junaid AM, Zafar SA, and Hamad AM. Automation Bias in Large Language Model–Assisted Diagnostic Reasoning among Physicians Trained in AI Literacy — A Randomized Clinical Trial. NEJM AI 2026 Apr 23;3(5):AIoa2501001.
13. Gaube S, Suresh H, Raue M, et al. Do as AI say: susceptibility in deployment of clinical decision-aids. npj Digital Medicine 2021;4.
14. Zarama V, Espinola JA, Camargo CA. Sepsis identification and bundle delivery in US emergency departments, 2012–2022: A nationally representative analysis. Journal of Critical Care. 2027 Feb 1;97:155740. doi:10.1016/j.jcrc.2026.155740
15. Rhee C, Train SE, Filbin MR, Park ST, Mohr NM, Zepeski A, et al. Complex Sepsis Presentations, SEP-1 Compliance, and Outcomes. JAMA Netw Open. 2025 Mar 19;8(3):e251100. doi:10.1001/jamanetworkopen.2025.1100
16. Fleck D, Gad H, Hogan Quigley B, Antar M, Sayed Ahmed A, Mahmoud MA, et al. The Effect of Clinical Ambiguity on the Decision-Making Process Among Intensive Care Unit Providers in Northern America Using Clinical Vignettes in Mixed Methods Study. Journal of Multidisciplinary Healthcare. 2025 May 31;18:3091–104. doi:10.2147/JMDH.S514274
17. Sivaraman V, Mason EP, Li ME, Tong J, King AJ, Kahn JM, et al. Intelligent Reasoning Cues: A Framework and Case Study of the Roles of AI Information in Complex Decisions. In: Proceedings of the 2026 CHI Conference on Human Factors in Computing Systems [Internet]. New York, NY, USA: Association for Computing

Machinery; 2026 [cited 2026 Jun 29]. p. 1–21. (CHI ’26). Available from: https://dl.acm.org/doi/10.1145/3772318.3790953doi:10.1145/3772318.3790953

18. Everett SS, Bunning BJ, Jain P, Lopez I, Agarwal A, Desai M, et al. From tool to teammate in a randomized controlled trial of clinician-AI collaborative workflows for diagnosis. npj Digit Med. 2026 Mar 18. doi:10.1038/s41746-026-02545-1
19. van Baalen S, Boon M, Verhoef P. From clinical decision support to clinical reasoning support systems. Journal of Evaluation in Clinical Practice. 2021 Jun 1;27(3):520–8. doi:10.1111/jep.13541
20. Dinc MT, Bardak AE, Bahar F, and Noronha C. Comparative analysis of large language models in clinical diagnosis: performance evaluation across common and complex medical cases. JAMIA Open 2025;8:ooaf055.
21. Festor P, Jia Y, Gordon AC, Faisal AA, Habli I, Komorowski M. Assuring the safety of AI-based clinical decision support systems: a case study of the AI Clinician for sepsis treatment. BMJ Health Care Inform. 2022 Jul 17;29(1). doi:10.1136/bmjhci-2022-100549 PubMed PMID: 10.1136/bmjhci-2022-100549.
22. Jeter R, Josef C, Shashikumar S, Nemati S. Does the “Artificial Intelligence Clinician” learn optimal treatment strategies for sepsis in intensive care? arXiv. 2019 Nov 22. doi:10.1038/s41591-018-0213-5 PubMed PMID: 30349085.
23. Altinger G, Maher CG, Jones CMP, et al. Multiple Suggested Care Alternatives and Decision-Making of Primary Care Physicians: A Randomized Clinical Trial. JAMA Network Open 2025;8:e2542949– e2542949.
24. Buçinca Z, Malaya MB, Gajos KZ. To Trust or to Think: Cognitive Forcing Functions Can Reduce Overreliance on AI in AI-assisted Decision-making. Proc ACM Hum-Comput Interact. 2021 Apr 22;5(CSCW1):188:1-188:21. doi:10.1145/3449287
25. Zhang S, Yu J, Xu X, Yin C, Lu Y, Yao B, et al. Rethinking Human-AI Collaboration in Complex Medical Decision Making: A Case Study in Sepsis Diagnosis. In: Proceedings of the 2024 CHI Conference on Human Factors in Computing Systems [Internet]. New York, NY, USA: Association for Computing Machinery; 2024 [cited 2024 Oct 22]. p. 1–18. (CHI ’24). Available from: https://dl.acm.org/doi/10.1145/3613904.3642343doi:10.1145/3613904.3642343

26. Martin-Loeches I, Singer M, Leone M. Sepsis: key insights, future directions, and immediate goals. A review and expert opinion. Intensive Care Med. 2024 Dec;50(12):2043–9. doi:10.1007/s00134-024-07694-z
27. Natali C, Marconi L, Dias Duran LD, Cabitza F. AI-induced Deskilling in Medicine: A Mixed-Method Review and Research Agenda for Healthcare and Beyond. Artif Intell Rev. 2025 Aug 27;58(11):356. doi:10.1007/s10462-025-11352-1
28. Kostick-Quenet KM, Gerke S. AI in the hands of imperfect users. npj Digit Med. 2022 Dec 28;5(1):197. doi:10.1038/s41746-022-00737-z
29. Ye Z, Chen Y, Huang X, Chen M, Wang W, Zhu H, et al. A five-phase evaluation framework for diagnostic and predictive medical artificial intelligence. npj Digit Med. 2026 Sep 3;9(1):678. doi:10.1038/s41746-026-03155-7
30. Gottesman O, Johansson F, Komorowski M, et al. Guidelines for reinforcement learning in healthcare. Nature Medicine 2019;25:16–8.
31. Agarwal N, Moehring A, Rajpurkar P, and Salz T. Combining Human Expertise with Artificial Intelligence: Experimental Evidence from Radiology. Working Paper 31422. National Bureau of Economic Research, 2023. doi: 10.3386/w31422. url: http://www.nber.org/papers/w31422.
32. Jacobs M, Pradier MF, McCoy TH, Perlis RH, Doshi-Velez F, and Gajos KZ. How machine-learning recommendations influence clinician treatment selections: the example of the antidepressant selection. Translational Psychiatry 2021;11.
33. Johnson A, Bulgarelli L, Pollard T, Horng S, Celi LA, and Mark R. MIMIC-IV (version 1.0). 2020.
34. Levin JM, Bukowski LA, Minson JA, and Kahn JM. The political polarization of COVID-19 treatments among physicians and laypeople in the United States. Proceedings of the National Academy of Sciences 2023;120. Publisher: Proceedings of the National Academy of Sciences.
35. Levin JM, Davis BS, Bukowski LA, Kahn JM. Willingness to treat with therapies of unknown effectiveness in severe COVID-19: a survey of intensivist physicians. Annals of the American Thoracic Society. 2022 Apr;19(4):633-9.

36. Malley BE, Levin JM, Kahn JM, Bukowski LA, Huang DT. Barriers to adopting evidence from Bayesian adaptive clinical trials in critical care: a randomized experiment in United States intensivists. CHEST Critical Care. 2026 Mar 7:100264.

## Acknowledgments

The authors thank Ali Smith and Arpit Mathur for assistance running the study, and Sivaraman Sivaswami, Mayank Goel, Haiyi Zhu, Jill Lehman, Rigney Turnham, and Julian Hong for critical feedback on the manuscript and analysis.

## Funding Statement

J.K. discloses support for this work from the National Heart, Lung, and Blood Institute (NHLBI) of the National Institutes of Health (NIH) under Award Number R35HL144804. NIH provided 100% of total project costs (approximately $125,000). The content is solely the responsibility of the authors and does not necessarily represent the official views of the National Institutes of Health. The remaining authors declare no other relevant funding.

## Author Contributions

V.S., A.K., J.K., and A.P. originated the study idea. V.S. led the model evaluation, interface and experimental design, data collection, analysis, and manuscript writing. V.S. and E.M. performed the qualitative coding and thematic analysis. V.T. provided guidance on the pre-registration and conducted the power analysis; J.F. assisted with analysis after data collection. J.L., S.S., J.F., A.K., J.K., and A.P. provided feedback throughout the research and contributed critical review and revisions to the manuscript.

## Competing Interests

The authors report the following competing interests: Dr. Saria reported holding a leadership position at Bayesian Health; receiving honoraria for lectures on artificial intelligence and health from various tech, biotech, healthtech, and professional societies; and holding equity in several companies, including Bayesian Health, Century Health, Specta Health, Midstream Health, Latent Health, and Duality Tech.

## Materials and Correspondence

Questions and requests for data may be directed to Adam Perer at adamperer@cmu.edu.

## Consent to Participate

Per our IRB-approved protocol, all human subjects consented to participation by reviewing the study's description, risks and benefits, compensation, data security practices, and investigator contact information on the online experiment platform, then clicking Begin. Participants could exit the study at any time.

## Clinical Trial Registration

This study was a randomized non-clinical trial that measured only simulated health outcomes on patient vignettes and was therefore not registered as a clinical trial. The study methodology was pre-registered at https://osf.io/h5urm.

Figure 1: AI outputs on a held-out patient cohort. (a) Empirical mortality risk estimates over the held-out patient cohort, visualized in a Uniform Manifold Approximation and Projection (UMAP) plot where each point represents a patient at a particular epoch. Different regions of the patient feature space take on different levels of mortality risk based on their most similar patients. (b) Estimated policy value difference of AI-recommended treatments compared to the original prescribing clinician policy in the held-out cohort, both overall and in subgroups based on case complexity. * - $p < 0.05$; ** - $p < 0.01$.

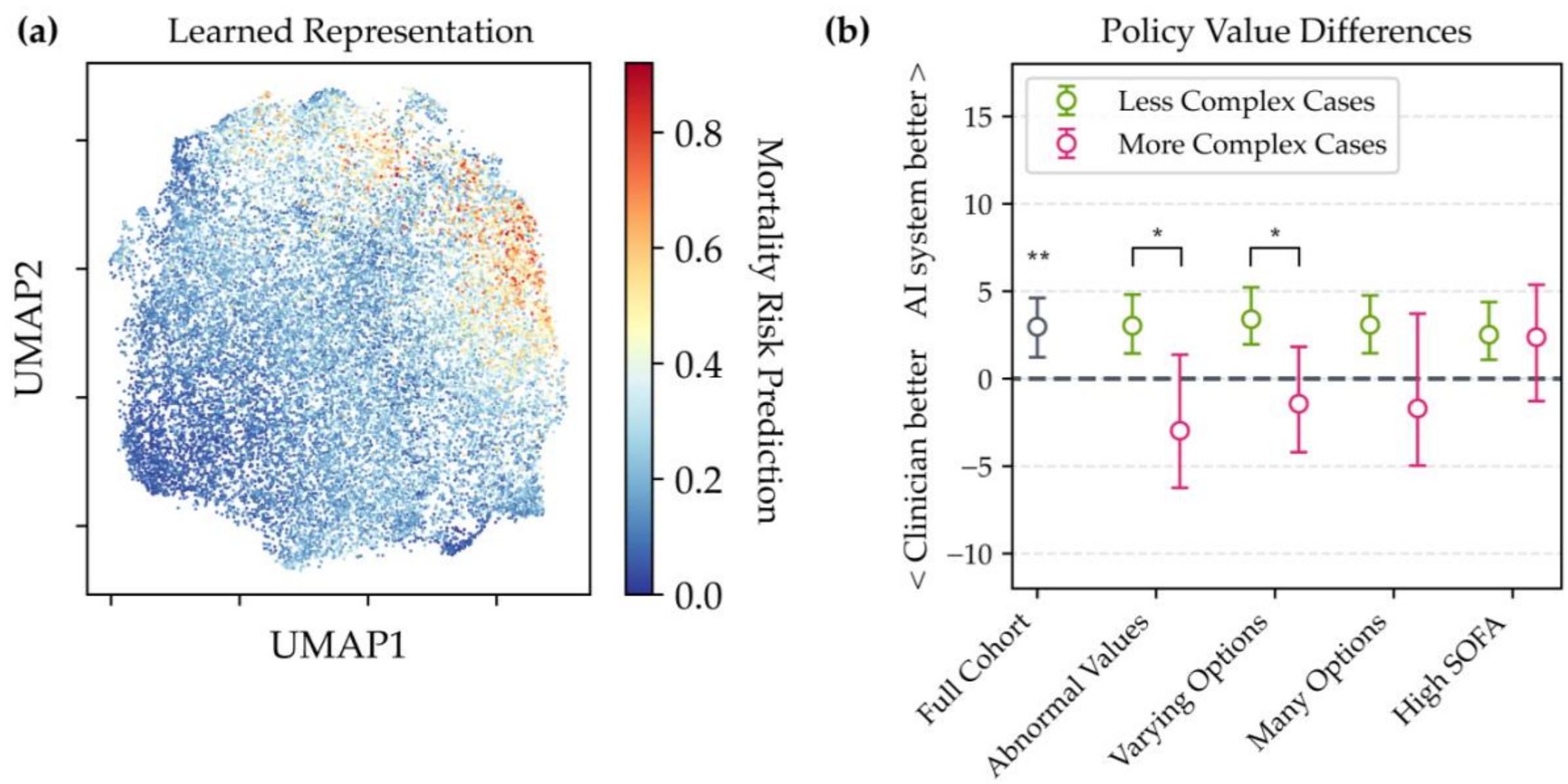

Table 1: Self-reported characteristics of the critical care physicians in our sample.

| **Characteristic** | **No. (%)** |
|---|---|
| Sex | |
| Female | 67 (29.4) |
| Male | 157 (68.9) |
| No response | 4 (1.8) |
| Race | |
| American Indian/Alaska Native | 0 (0) |
| Asian | 60 (26.3) |
| Black | 0 (0) |
| Multiracial | 5 (2.2) |
| Native Hawaiian/Pacific Islander | 2 (0.9) |
| White | 147 (64.5) |
| No response | 14 (6.1) |
| Ethnicity | |
| Hispanic or Latino | 5 (2.2) |
| Not Hispanic or Latino | 124 (54.4) |
| No response | 99 (43.4) |
| Base specialty | |
| Anesthesiology | 15 (6.6) |
| Emergency medicine | 14 (6.1) |
| Internal medicine, not pulmonary | 16 (7.0) |
| Internal medicine, pulmonary | 160 (70.2) |
| Neurology | 1 (0.4) |
| Surgery | 8 (3.5) |
| Other | 14 (6.1) |
| Practice setting | |
| Community | 27 (11.8) |
| Academic, not university affiliated | 24 (10.5) |
| Academic, university affiliated | 175 (76.8) |
| Other | 2 (0.9) |
| Clinical time | |
| More than none but less than 50% | 87 (38.2) |
| 50 - 95% | 76 (33.3) |
| 95% or more | 62 (27.2) |
| No response | 3 (1.3) |
| Self-rated sepsis treatment expertise | |
| Expert | 178 (78.0) |
| Proficient | 39 (17.1) |
| Competent | 10 (4.4) |
| Beginner | 0 (0) |
| Novice | 1 (0.4) |
| Experience in intensive care unit (ICU) | |
| > 10 years | 105 (46.1) |
| 5 - 10 years | 105 (46.1) |
| 3 - 5 years | 18 (7.9) |
| 1 - 2 years | 0 (0) |
| < 1 year | 0 (0) |

Figure 2: Decision quality under observed and simulated decision-making patterns. Blue bars indicate the observed rate of choosing optimal treatments; gray bars indicate the simulated optimal treatment rate if clinicians always agreed with the AI advice. Error bars indicate 95% confidence intervals; ** - $p < 0.01$.

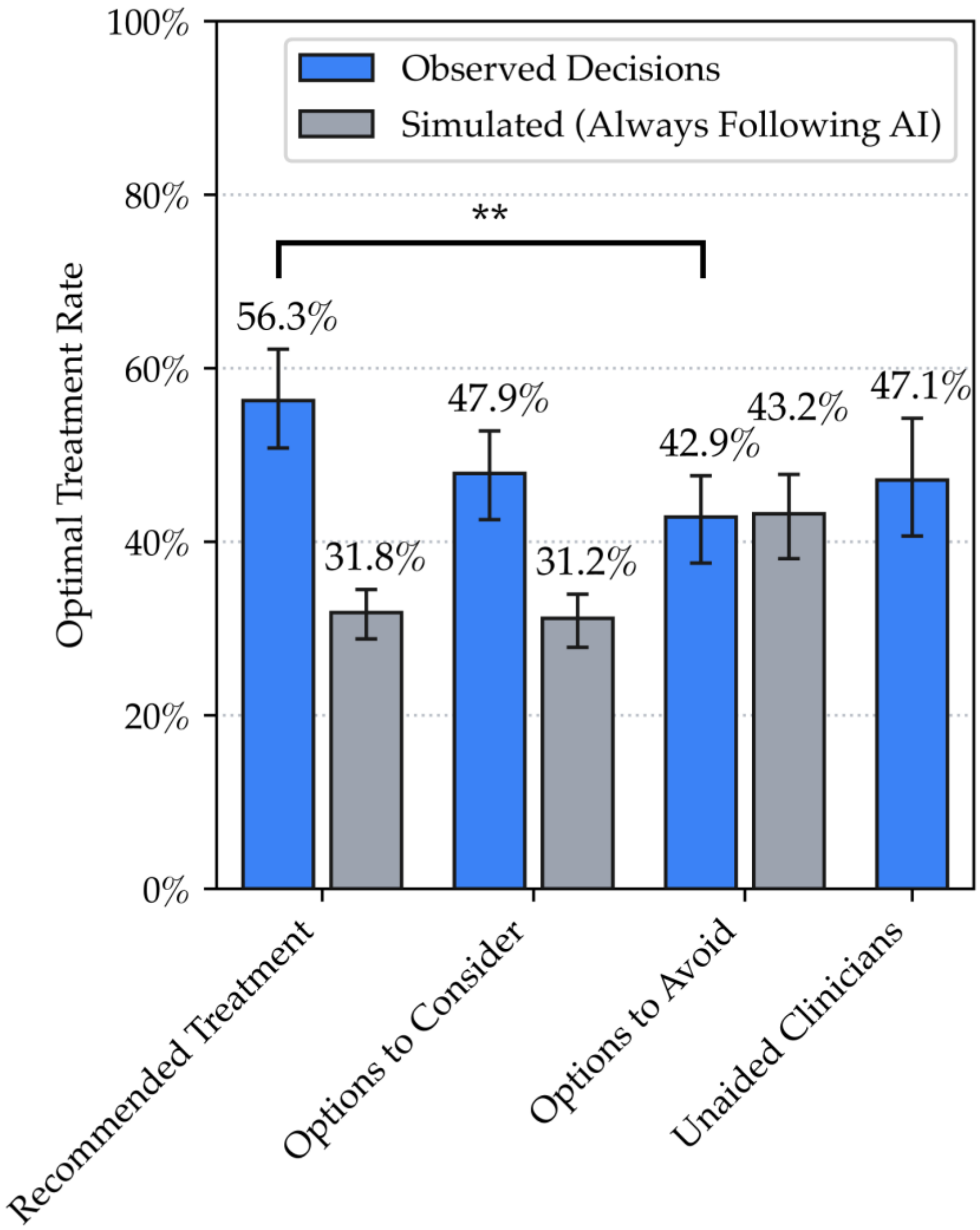

Figure 3: (a-c) Top row: Changes in plan alignment with AI advice between participants in each AI condition and unaided clinicians. For example, a positive value for Mentioned by AI indicates that participants who saw the AI selected AI-mentioned treatments more often than unaided clinicians. (a-c) Bottom row: Rates at which taking Mentioned, Adjacent, and Other plans resulted in an optimal treatment. (d) An example of the 2-dimensional treatment grid on which Mentioned, Adjacent, and Other decisions were defined; specific assignments depend on the patient case and AI condition. Cell shading and numbers indicate expert-rated quality scores, with the highest value on the grid representing the optimal decision. Error bars indicate 95% CIs; * - $p < 0.05$; ** - $p < 0.01$; *** - $p < 0.001$. IVF - intravenous fluids; VP - vasopressors; Diu. - diuretics; cond. - vasopressor decision conditional on response to fluid.

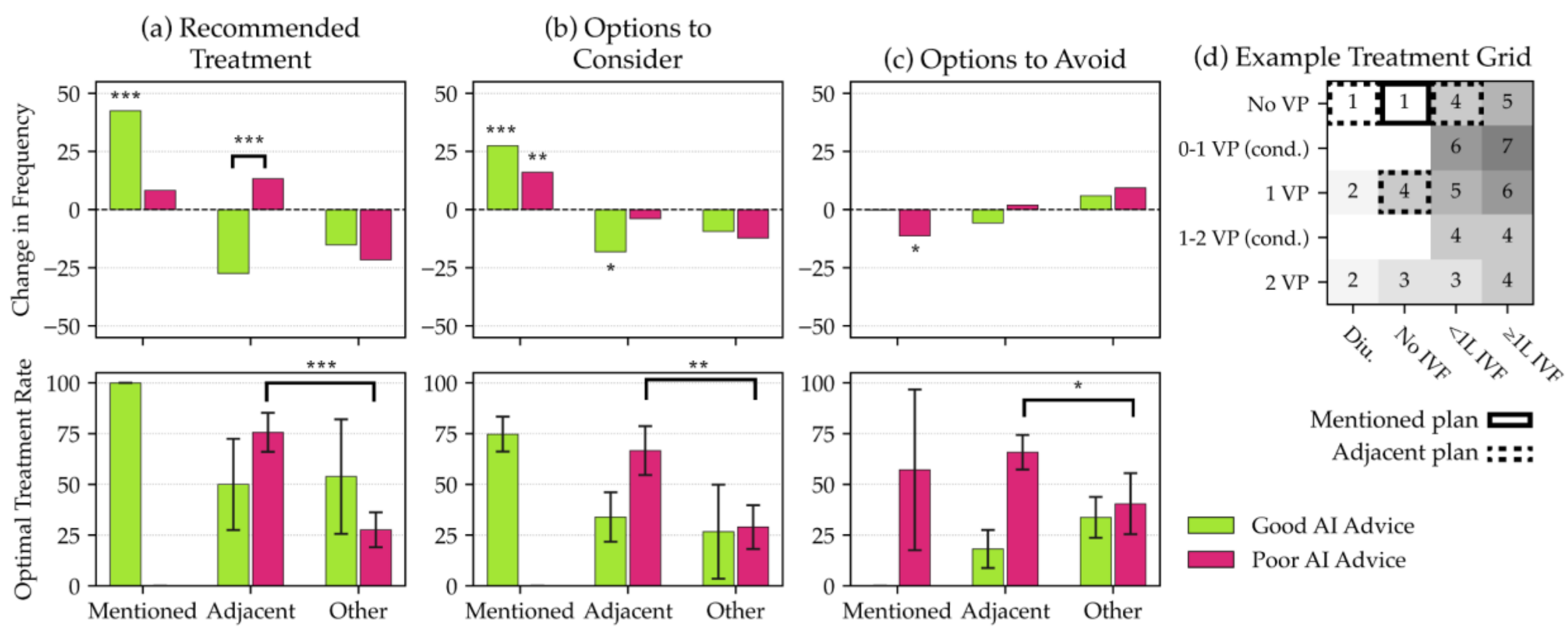

Figure 4: Participant flowchart.

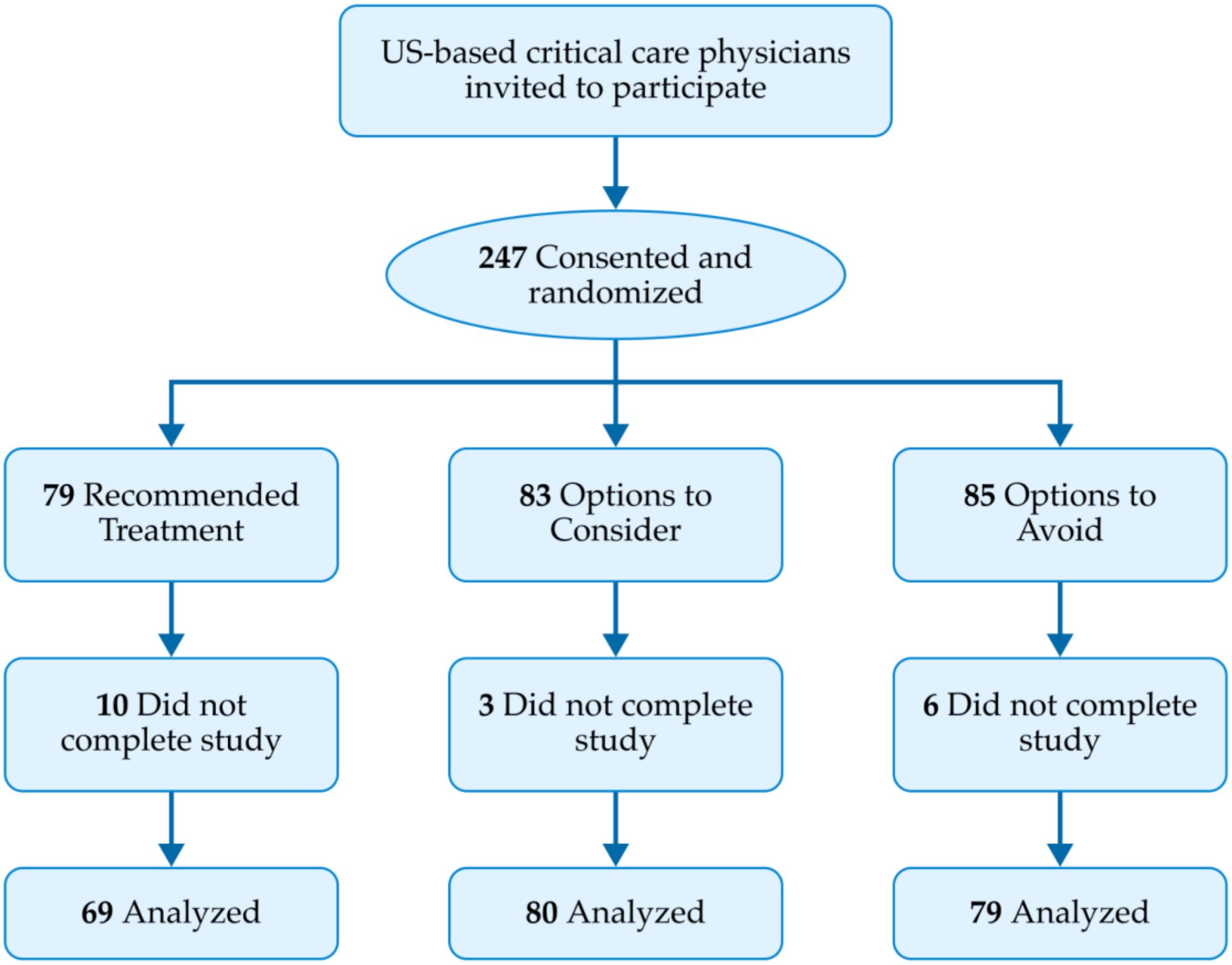

Figure 5: AI information types tested in this experiment.

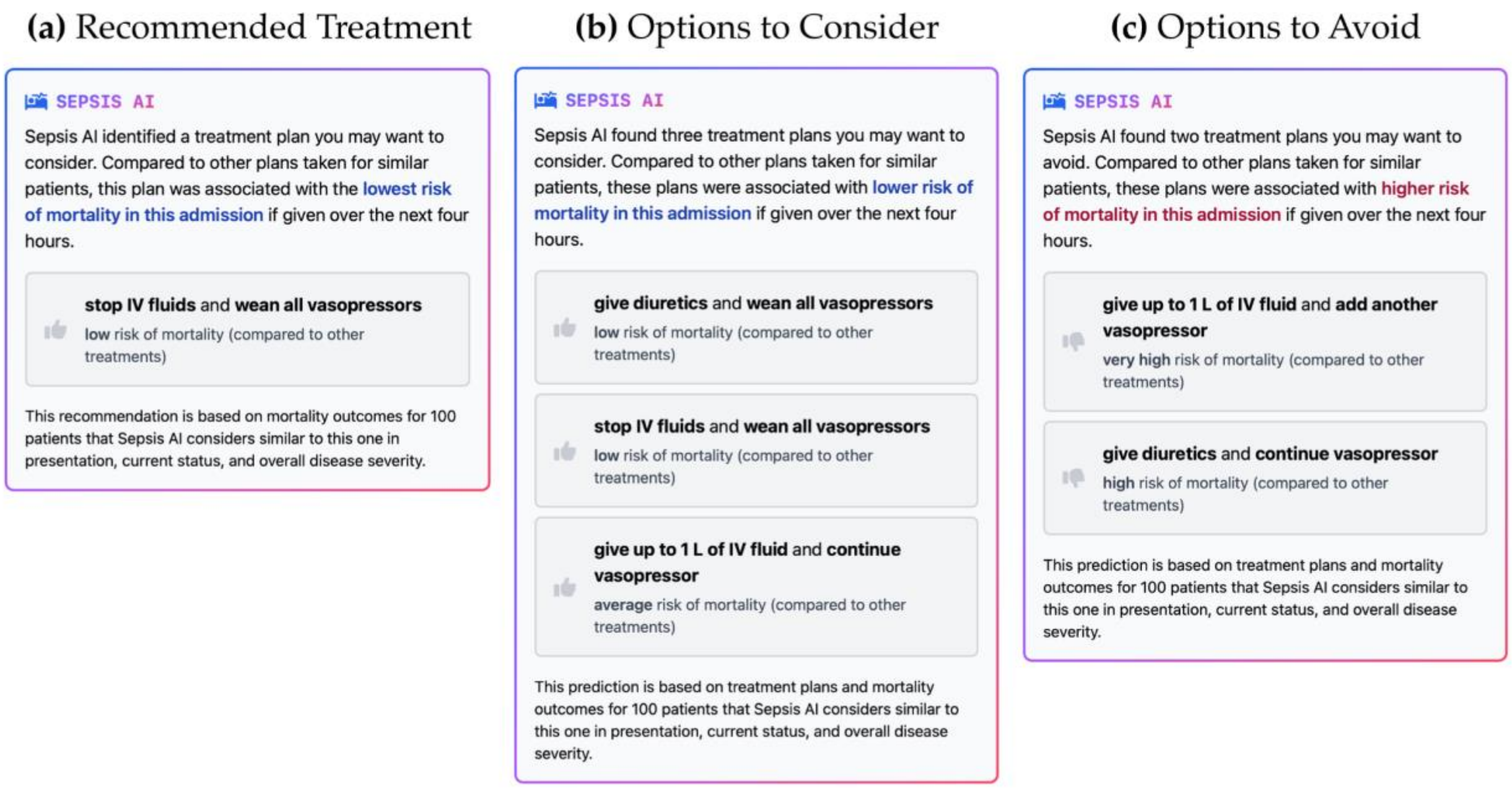

## Supplementary Information

## S.1 AI System Details

### S.1.1 Data Preparation and Model Training

The model in this study was a transformer encoder model previously developed by Sivaraman et al.[1] using patient data from the publicly-available MIMIC-IV dataset[2]. Patients were included if they were in the intensive care unit (ICU) for at least 12 hours and had possible sepsis, defined as microbial cultures taken and antibiotics prescribed within 24 hours of each other. Possible sepsis was defined broadly with the intention of maximizing the amount of data available for model training (see below). Data was excluded for patients who were in the ICU longer than the 95th percentile of ICU stay length (about 17 days) and where comfort measures only (CMO) orders were in place. Patient data was aggregated at four-hour intervals to minimize data missingness, consistent with prior work[3]. These transformations resulted in a cohort of 30,398 patients with an average of 30 timesteps per patient; half of these patients were used for model training while the other half were used for validation and case selection.

The model was trained using unsupervised learning, a technique often used to identify patterns of similarity in high-dimensional data without requiring labels. Specifically, it was optimized to perform a “denoising” task in which the model was given corrupted versions of patient data (with values shifted or masked at random) and trained to output the original values. The inputs to the model included demographics, comorbidities, charted observations, medications, and prior hemodynamic management decisions for each patient-timestep.

After training, the model was used to generate vector representations of all patients in the dataset, allowing us to retrieve data on patient-timesteps with similar representations. These “nearest neighbors” were stratified by Sequential Organ Failure Assessment (SOFA) scores, a widely-used measure of sepsis disease severity; this ensured that a given patient's recommendations would be based on patients at a similar degree of illness. Among the 100 nearest-neighbor cases for a given patient-timestep, we counted how often similar patients received each of the twelve different treatment plans, and we estimated mortality risk for each patient-timestep as the empirical rate of in-hospital mortality among its nearest neighbors. We verified that the overall mortality risk predictions were fairly accurate on the evaluation set, achieving an AUROC score of 0.742 without any fine-tuning specific to the task.

### S.1.2 Retrospective Evaluation Details

The retrospective evaluation used Weighted Importance Sampling (WIS), a technique to estimate the expected value of a treatment policy using data collected under a different policy. It works by assigning each epoch j within each trajectory i an importance score based on the likelihood of choosing the true treatment under both the evaluation policy $\pi_1$ and the behavior policy $\pi_0$, accumulated over all previous epochs:

$$\rho_i^j(\pi_0, \pi_1) = \frac{\prod_{k=0}^{j} \pi_1(a_i^k | x_i^k)}{\prod_{k=0}^{j} \pi_0(a_i^k | x_i^k)}$$

The likelihood under the AI policy was defined to be $1 - \epsilon$ if $a_i^j$ was equal to the recommended action, and $\frac{\epsilon}{|A|-1}$ otherwise ($|A|$ represents the number of possible actions). The likelihood under the clinician's policy, meanwhile, was estimated by a simple classifier trained on the original model's inputs (this study used a multi-class XGBoost model). Intuitively, the importance score would be larger when the AI policy is aligned with the true action throughout a trajectory, upweighting such trajectories in the final estimate. To mitigate the effect of very large weights from a few trajectories dominating the WIS estimates, the top 0.1% of importance weights were clipped to the 99.9th percentile weight value.

WIS accumulates the weighted sum of "rewards" obtained over each trajectory. We defined the rewards at each epoch identically to previous sepsis treatment work:[4]

$$r_i^j = 15S_i \times I(j = |X_i|) - 0.025I(SOFA_i^{j+1} = SOFA_i^j \wedge SOFA_i^j > 0) - 0.125(SOFA_i^{j+1} - SOFA_i^j)$$

where $S_i$ is set to –1 if the patient's discharge outcome was mortality and +1 otherwise, and $|X_i|$ is the length of trajectory i. This reward function heavily penalizes mortality outcomes at the end of each trajectory, while adding intermediate rewards for changes in the patient's disease severity measured by SOFA score.

Given this information, the WIS estimator for the value of a treatment policy is given by

$$\mu_{WIS}(\pi_0, \pi_1) = \frac{1}{\sum_{i=1}^{N} \sum_{j=1}^{|X_i|} \rho_i^j(\pi_0, \pi_1)} \sum_{i=1}^{N} \sum_{j=1}^{|X_i|} \rho_i^j(\pi_0, \pi_1)\, G_i^j,$$

where $G_i^j = \sum_{k=j}^{|X_i|} \gamma^{k-j} r_i^j$ represents a discounted sum of rewards over the remainder of the trajectory ($\gamma < 1$ is a discount factor). We computed the policy value of the AI's and the clinicians' policies as $\mu_{WIS}(\pi_{clin}, \pi_{AI})$ and $\mu_{WIS}(\pi_{clin}, \pi_{clin})$, respectively, then reported the difference between the two estimates across 10,000 bootstrapped replications.

For each of the four measures of complexity, we obtained complex case subgroups by selecting patients for whom each criterion was true in at least 20% of epochs. As shown in Table S1, these subgroups ranged from 15% to 40% of the total evaluation cohort. The mortality prediction performance (a proxy for model representation quality) decreased in all subgroups except for Many Options, with the greatest degradation in High SOFA cases (a 6 percentage point decrease). Also, the agreement between the AI recommendations and the original prescribing clinician actions was lower than the overall cohort across all subgroups, suggesting that the system would be more often discordant with clinicians' typical practices in complex cases.

Table S1: Performance metrics for complex patient subgroups.

| **Subgroup** | **Number of Patients (%)** | **Mortality Prediction AUROC** | **Agreement Between Recommendation and Original Clinician** | |
|---|---|---|---|---|
| | | | **Fluids** | **Vasopressors** |
| All | 6,928 (100%) | 0.742 | 67.6% | 90.5% |
| Abnormal Values | 1,465 (21.1%) | 0.729 | 64.0% | 84.1% |
| High SOFA | 1,819 (26.3%) | 0.679 | 67.1% | 83.6% |
| Many Options | 1,104 (15.9%) | 0.746 | 58.6% | 74.6% |
| Varying Options | 2,767 (39.9%) | 0.731 | 61.0% | 81.8% |

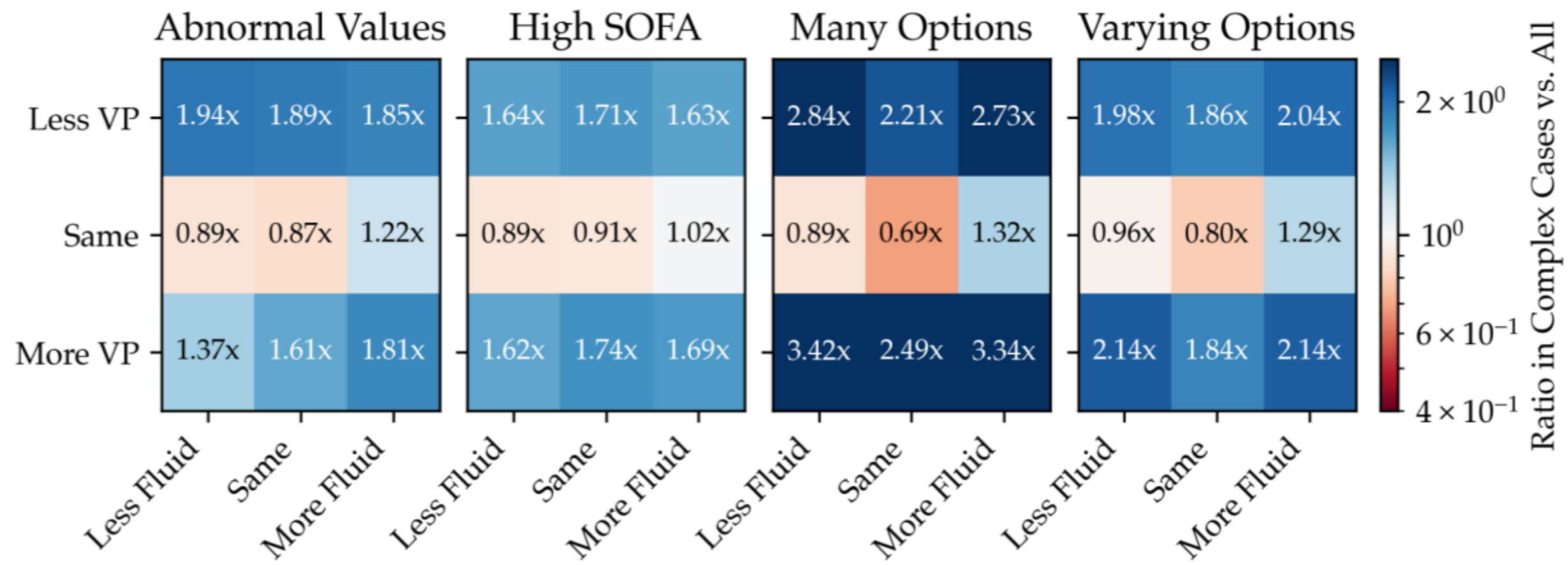

Figure S1: Changes in AI recommendation relative to true clinician action in complex case subgroups. Overall, the model was more likely to recommend different treatments than the original prescribing clinician in complex cases, and particularly more likely to recommend different vasopressor plans.

### S.2 Case-Level Expert Ratings and Results

This section provides further detail on the AI outputs, decision quality scores, and shifts in treatment choice frequencies at the patient case level. As shown in Figure S2, each case was associated with between 12 and 14 possible treatment decisions spanning a two-dimensional choice grid of intravascular volume (x-axis) and vasopressors (y-axis).

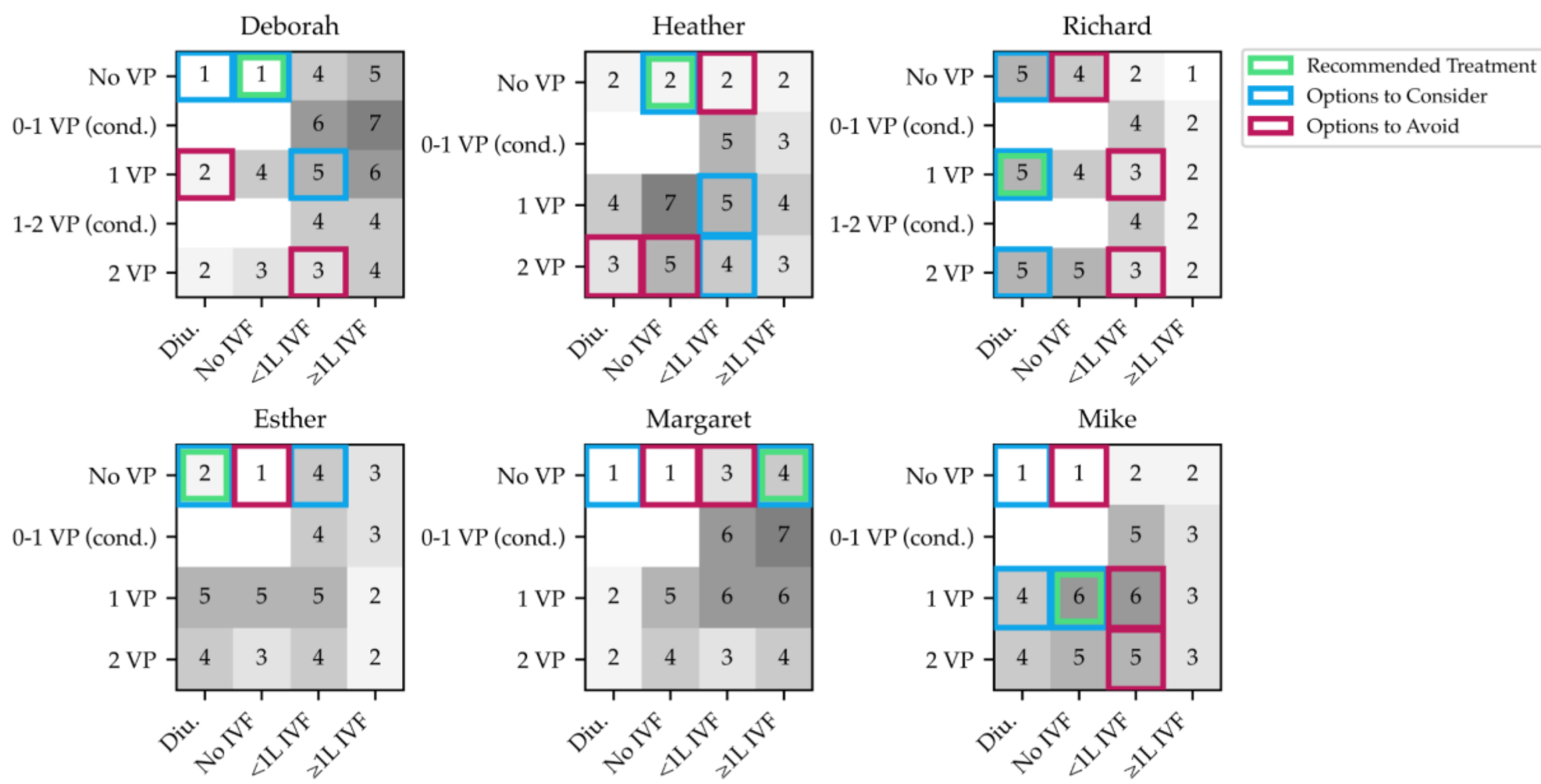

Figure S2: Consensus quality scores from the expert panel, displayed on the grid of possible treatment choices for each case. Higher scores signify treatments more likely to help the patient. The colored boxes indicate plans that were mentioned by each of the AI conditions. “Diu.” - diuretics; “IVF” - intravenous fluid; “VP” - vasopressor; “cond.” - treatment conditional on response to fluid.

**Expert Panel Ratings.** The expert panel rated each decision for each case on a 7-point Likert scale where 1 indicated most likely to harm the patient and 7 indicated most likely to help the patient. In treatments where the panelists’ individual ratings differed by more than three quality rating points, the final quality score was assigned by consensus after discussion; the remaining treatments were assigned the median quality score rating. A total of 28 out of 88 total treatment decisions were discussed, with the vignettes for Deborah, Richard, and Esther requiring the greatest amount of deliberation. The final quality scores spanned the entire 7-point spectrum, with 29.5% of treatments rated as at least somewhat likely to help the patient (score of 5 or higher). The maximum score for each case ranged from 5 to 7, indicating that there was not always a treatment that would be certain to benefit the patient.

**AI-Mentioned Treatments.** The specific treatments mentioned by each AI condition for each case are shown with colored boxes in Figure S2. Each system mentioned between 1 and 3 treatments per case; note that the plan mentioned by Recommended Treatment is always one of those mentioned by Options to Consider, and none of the AI systems were able to surface a conditional plan because they were not part of the modeled treatment action space.

**Changes in Decisions with AI vs. Unaided Clinicians.** Figure S3 shows the shifts in treatment choice frequency in each AI condition compared to unaided clinicians; AI-mentioned plans are shown by black boxes. In the Recommended Treatment condition (left column), we observe that the AI-mentioned plan is always associated with an increase in frequency compared to unaided clinicians. Notably, plans close to but not identical to the mentioned plan (Adjacent plans) also increase in frequency. For example, in Deborah's case the choice to add a vasopressor while avoiding fluids is associated with an 18pp increase, and in Margaret's case the conditional vasopressor choice increases in frequency by 24pp. In Options to Consider (middle column) there is always at least one mentioned plan that increases in frequency; however, there is less evidence of clinicians choosing Adjacent plans. Finally, in Options to Avoid, we observe that the initial frequency of the mentioned treatments is low across all cases except Mike's, explaining this condition's limited effects.

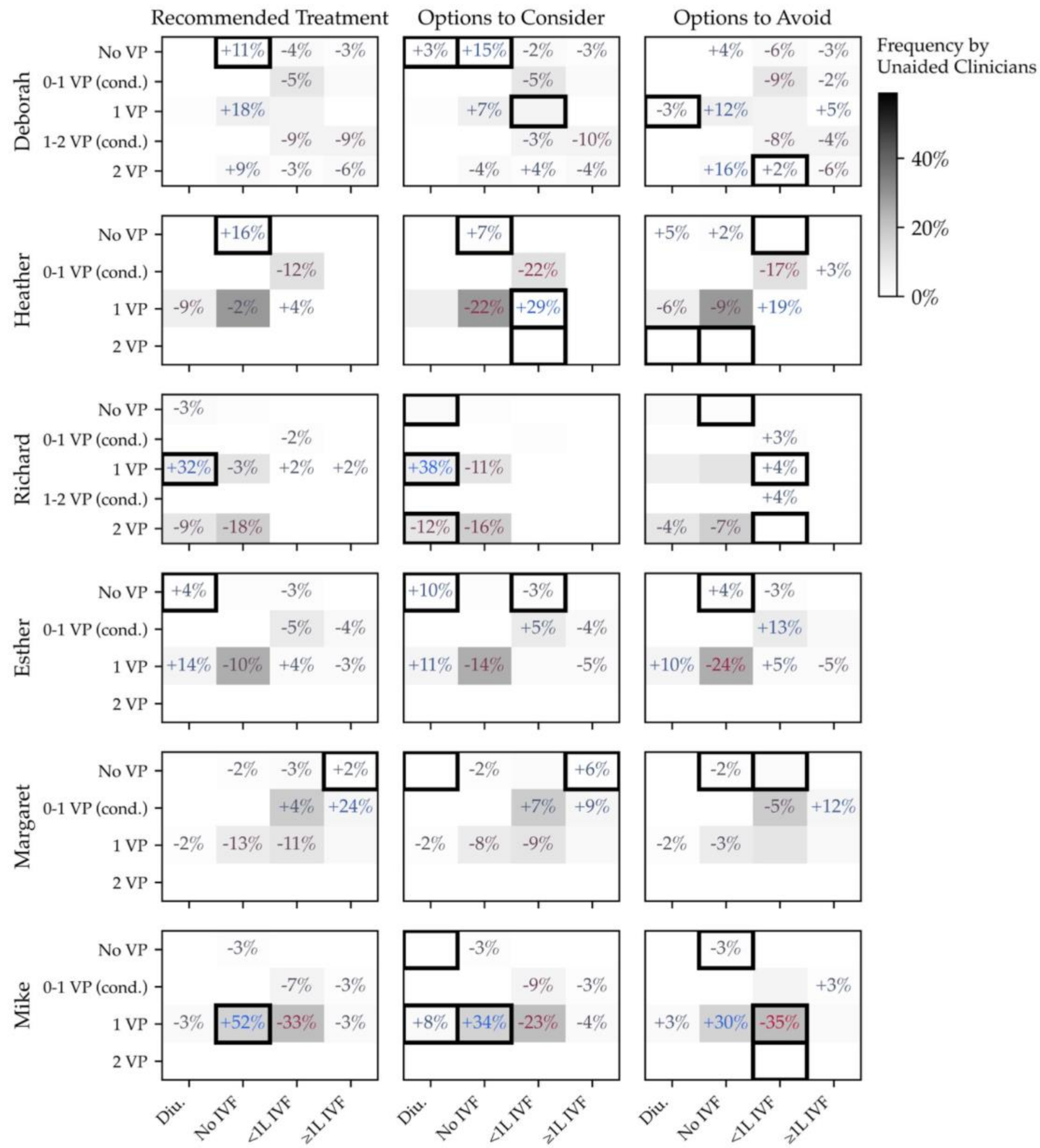

Figure S3: Changes in case-level decisions within each AI condition. The text in each cell shows the absolute percentage difference between clinicians who saw the AI for the case and unaided clinicians. Cell shading indicates absolute frequency of decisions by unaided clinicians; bordered boxes indicate plans mentioned by each AI condition. “Diu.” - diuretics; “IVF” - intravenous fluid; “VP” - vasopressor; “cond.” - treatment conditional on response to fluid.

### S.3 Additional Protocol Details

The pre-registered study protocol is provided at https://osf.io/h5urm/overview. Additional information required by the CONSORT reporting standard is given below.

**Patient and Public Involvement.** Patients and the public were not involved in the design, conduct, or reporting of the trial. Critical care physicians unaffiliated with the project informed the design of the study through prior qualitative studies[1,5].

**Changes to Trial Protocol.** No changes were made to the recruiting and participation methods. As described in Supplementary Sec. S.4.1, the originally pre-specified analysis used ordinal regression on the expert-assigned Likert quality scores. This analysis showed similar significance trends to the rate of selecting optimal treatments but is more complex to interpret; therefore, it is provided in Supplementary Sec. S.4.1 alongside other variants. The exploratory analysis of alignment with AI advice was added after data collection was completed to explain the findings on the primary outcome.

**Identification of Harms.** The study was determined to pose minimal risk of harm to participants by the University of Pittsburgh IRB.

## S.4 Additional Statistical Analyses

### S.4.1 Alternative Decision Quality Analyses

**Ordinal Regression Analysis.** The pre-registration for this study specified the use of mixed-effects ordinal regression on the 7-point Likert scale ratings for each decision. For ease of interpretation, this analysis was replaced with logistic regression where the outcome was whether participants chose a treatment with the highest possible quality score for each case (Sec. 3.1). The originally pre-registered model included the same terms as the one presented in the main text, with the AI information condition as a fixed effect and patient case and respondent ID as random effects. Odds ratios (ORs) below indicate the likelihood of being in the next higher quality score category between two conditions. p-values for post-hoc pairwise comparisons were corrected with the Tukey method. The distributions of decision quality scores for clinicians in each AI condition is shown in the blue bars in Figure S4.

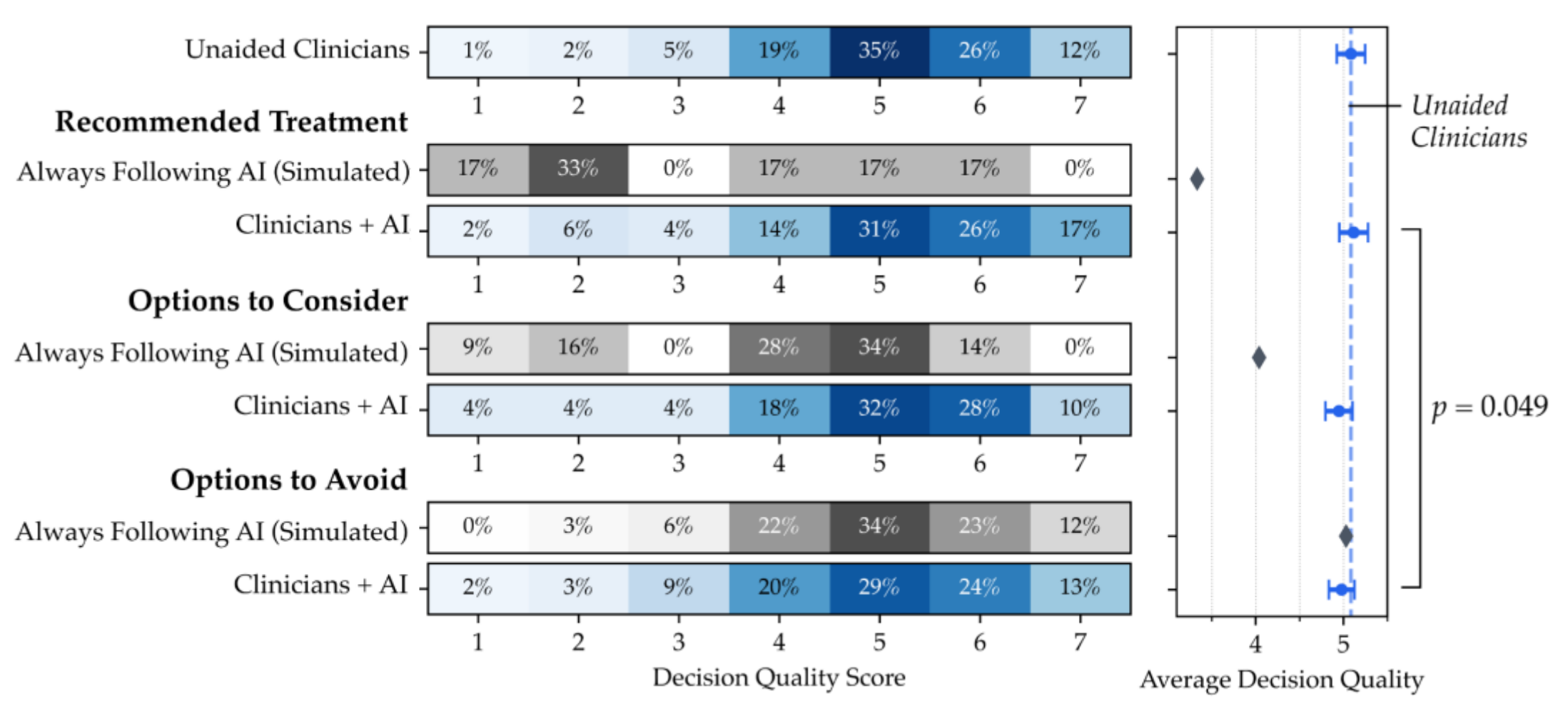

Figure S4: Distribution of decision quality scores in each AI condition. Gray rows show expected decision quality if clinicians always followed AI advice, while blue rows show observed decision quality scores. The right panel shows the same decision quality scores in aggregate, with error bars indicating 95% confidence intervals (CIs) around the observed means.

The three AI information types had significantly different coefficients in this model ($p$ = 0.02, Wald $\chi^2$ = 7.83, 2 df), indicating that AI information type affected decision quality differently. In pairwise comparisons of the four conditions, Recommended Treatment resulted in significantly higher decision quality than Options to Avoid ($p$ = 0.049, OR = 1.46 [1.10,1.95]). As in the analysis presented in the main text, there were no other statistically significant differences in pairwise comparisons between the AI conditions after Tukey correction. Recommended Treatment had directionally higher decision quality compared to unaided clinicians (OR = 1.22 [0.89,1.67]), while Options to Consider and Options to Avoid had directionally lower decision quality (OR = 0.87 [0.64,1.18] and OR = 0.83 [0.62,1.13], respectively). This analysis did not statistically evaluate or compare the theoretical decision quality scores of always following the AI advice.

**Alternate Threshold for Optimal Choices.** We also tested a version of the analysis where instead of designating only the highest-scoring treatments according to the expert panel as "optimal," we chose treatments that a clinician might consider "good enough" (i.e., either the best possible score for the case or at least a score of 6 out of 7, indicating that the treatment would likely help the patient). The results for this analysis were very similar to those presented in the main text. The three AI information types had significantly different coefficients in this model ($p$ = 0.0039, Wald $\chi2$ = 11.1, 2 df). In pairwise comparisons of the four conditions, Recommended Treatment resulted in significantly more frequent "good-enough" decisions than Options to Avoid ($p$ = 0.005, OR = 1.89 [1.30,2.75]); no other pairs of conditions had statistically significant differences. Compared to unaided clinicians, participants in Recommended Treatment made good-enough decisions directionally more often (OR = 1.47 [0.98,2.20]), those in Options to Consider made good-enough decisions at roughly the same rate (OR = 1.04 [0.71,1.54]), and those in Options to Avoid made them directionally less often (OR = 0.77 [0.53,1.14]). In the model adjusting for the quality of the AI advice, Recommended Treatment once again led to significantly more frequent good-enough decisions than unaided clinicians ($p$ = 0.008, OR = 2.26 [1.36,3.74]), those in Options to Consider ($p$ =

0.042, OR = 1.74 [1.15,2.63]), and those in Options to Avoid ($p < 0.001$, OR = 2.93 [1.81,4.75]).

### S.4.2 Alignment in Options to Consider and Options to Avoid

The following results provide more detail on the secondary analysis of alignment with AI advice in the main text.

In the Options to Consider condition (Fig. 3(b)), the pattern of participants' alignment with AI advice was less dependent on advice quality (3/6 cases each with good and poor advice). Participants in Options to Consider chose one of the plans mentioned by the AI 22.9% more often than unaided clinicians (15.6% to 38.5%, $p < 0.001$), with similar trends for both good (29.7% to 57.2%, $p < 0.001$) and poor advice (1.8% to 17.8%, $p = 0.005$). The degree of shift was not significantly different by advice quality. They selected Adjacent plans directionally less often overall (47.1% to 36.1%, $p = 0.058$), with a significant decrease when the advice was good (52.3% to 34.1%, $p = 0.024$) and minimal change when it was poor (42.1% to 38.2%, $p > 0.99$). This degree of shift was not significantly different in magnitude. As shown in the right panel of Fig. 2(b), Adjacent decisions for this condition were higher-quality than Other decisions (OR = 3.68 [1.61,8.41], $p = 0.01$) when the advice was poor and not significantly different when advice was good, a trend similar to Recommended Treatment.

The effect of Options to Avoid (Fig. 3(c)) was relatively small compared to the other two conditions, both when it contained good advice (3/6 cases) and poor advice (3/6 cases). Participants overall chose Mentioned plans directionally less often than unaided clinicians in this condition (10.2% to 4.6%, $p = 0.073$), as one would expect given the Mentioned plans in this condition were high-risk. In cases with good advice quality, participants did not choose either Mentioned or Adjacent plans at a significantly different rate than unaided clinicians, likely because participants already chose Mentioned plans very infrequently (Mentioned: 5.2% to 5.0%, $p > 0.99$; Adjacent: 47.0% to 41.2%, p >

0.99). When this AI gave poor advice (i.e. recommended avoiding a high-quality plan), participants did significantly shift away from the Mentioned plans (15.5% to 4.1%, $p = 0.015$), indicating that participants adjusted to accommodate the AI advice. The magnitude of the shifts was not significantly different by advice quality. There was also no corresponding significant shift towards Adjacent plans (69.1% to 71.0%, $p > 0.99$). Similar to the other conditions, Adjacent plans were significantly more likely to be optimal than Other plans when advice was poor (OR = 2.90 [1.36,6.20], $p = 0.024$) and not significantly different when advice was good. However, because participants' behavior changed little whether its advice was good or bad, this condition had minimal impact.

### S.4.3 Analysis of Usability Scores

We collected usability ratings for each AI condition by asking participants at the end of the survey to provide ratings across five scales adapted from the Technology Acceptance Model 3 (TAM-3). We conducted a prespecified one-way analysis of variance (ANOVA) of participants' overall usability rating, as measured by the sum of their ratings for the five scales. There were no significant differences in perceived usability between the three AI interfaces ($F(2,225) = 1.30$, $p = 0.275$); individual constructs did not show significant differences by AI condition either (distributions of ratings shown in Figure S5). This finding suggests that the differences in how participants used the AI were not simply due to differences in usability.

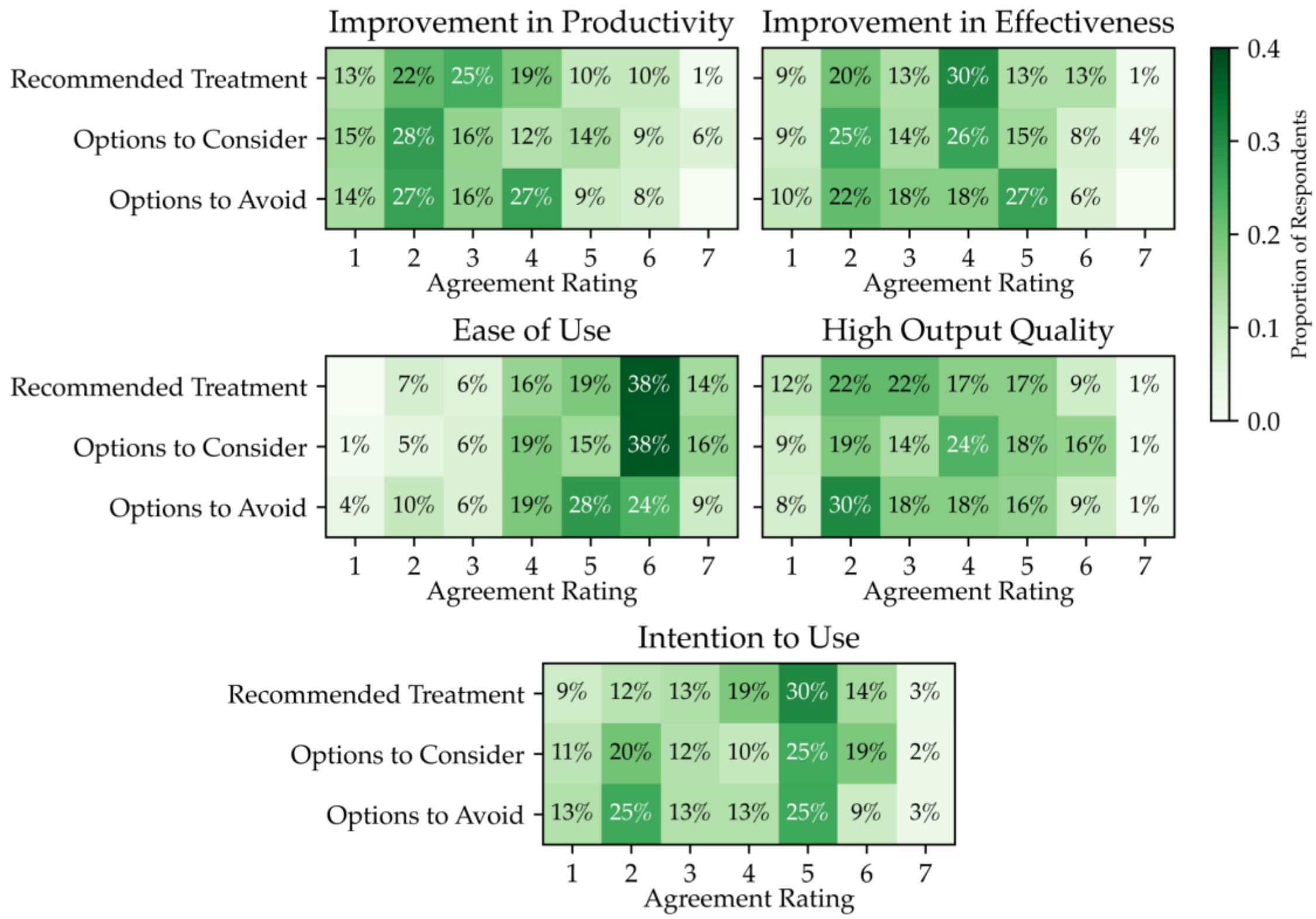

Figure S5: Distribution of Likert-scale ratings for post-task usability questions; higher agreement ratings indicate greater perception of usability.

### S.4.4 Self-Reported Influences of AI

After each case, participants were asked whether the AI influenced their thinking in eight specific patterns. These self-reported influences were analyzed with logistic regression on each influence pattern, with AI advice type as a fixed effect (p-values corrected with the Holm method). Figure S6 shows the proportion of participants that selected each of the eight pre-defined influence patterns in at least one case, ordered by the effect of AI condition. The influence patterns that differed most by AI condition were giving participants an initial plan ($p = 0.019$), prompting them to second-guess their plan ($p = 0.019$), and prompting to re-assess the patient ($p = 0.16$). Participants who received the Recommended Treatment were more likely to endorse all three of these influence patterns compared to those in other conditions. These findings suggest that those who saw the Recommended Treatment generally perceived more influences of the AI on their thinking, both in determining their initial decision and causing them to question their assessment.

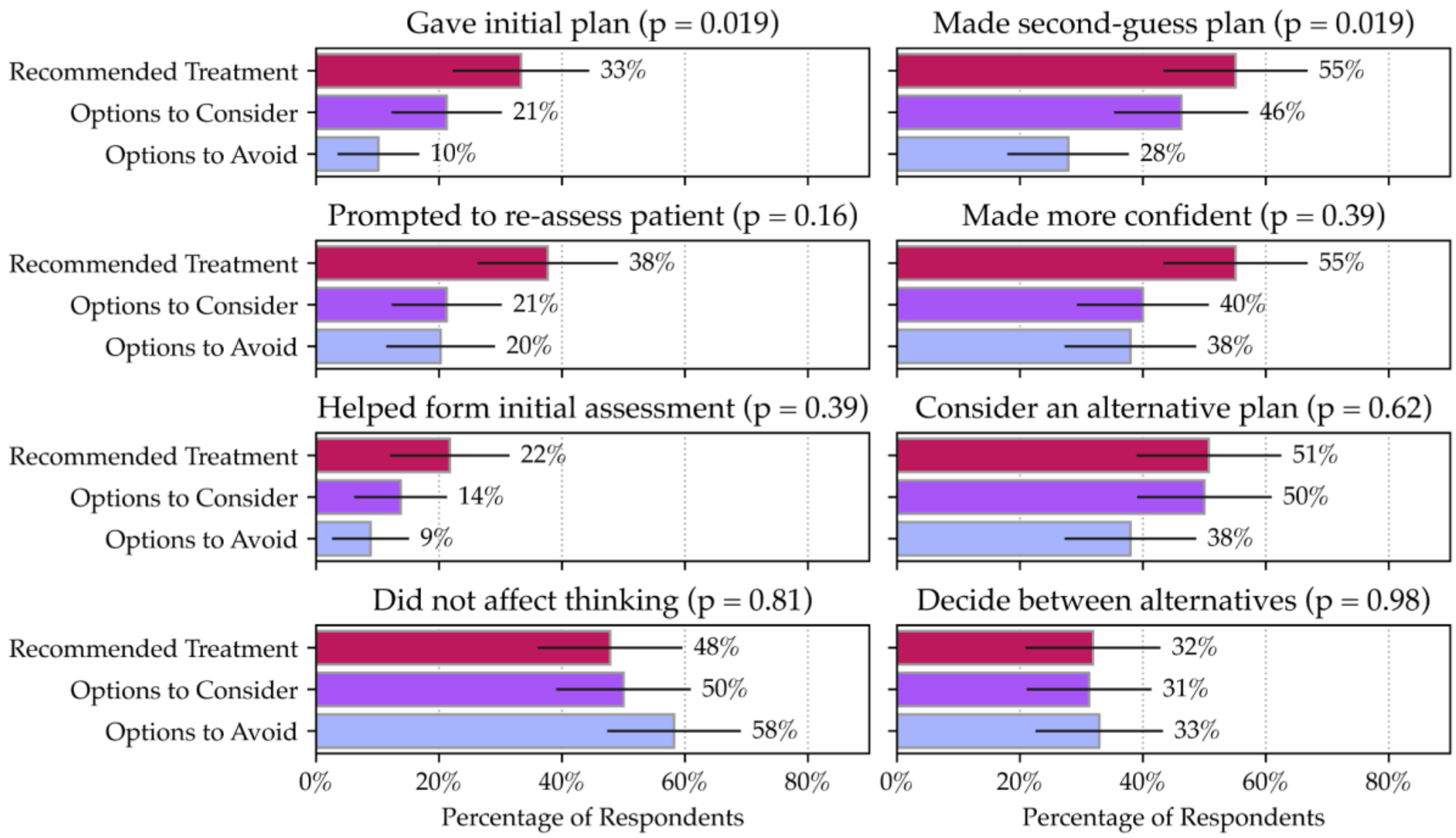

Figure S6: Percentages of respondents in each condition who selected various patterns of influence of the AI on their reasoning in at least one case. Error bars represent 95% confidence intervals around each proportion; p-values reflect the impact of AI advice type on the rate of positive responses.

### S.4.5 Qualitative Thematic Analysis

All themes identified from participants' open-ended responses are listed in Table S2. The most common themes were users' requests for a causal explanation of AI advice (23 participants), dispreference for Options to Avoid (23 participants), and an observation that additional contextual details such as fluid status and bedside assessments would help drive decisions (22 participants). Across conditions, participants found the AI-CDSS helpful as a prompt to reconsider their treatment plan (12 participants) and to catch potential errors (11 participants). They also consistently expressed desires for improved AI advice quality in these complex cases (19 participants), justifications from the clinical literature (13 participants), and more specific recommendations (12 participants).

Table S2: Full list of themes derived from participants' open-ended responses, along with counts of participants in each condition who expressed them. RT = Recommended Treatment; OC = Options to Consider; OA = Options to Avoid.

| Theme | Total | RT | OC | OA | Theme | Total | RT | OC | OA |
|---|---|---|---|---|---|---|---|---|---|
| Desire for a Causal Explanation for AI Advice | 23 | 7 | 8 | 8 | AI Potentially Useful for Medication Selection | 5 | 1 | 1 | 3 |
| What Not to Do Less Helpful than What to Do | 23 | 0 | 0 | 23 | Potential for AI to Help with Continuous Evaluation | 4 | 1 | 2 | 1 |
| Rely on Contextual Information | 22 | 6 | 5 | 11 | Helpful to Give Initial Plan | 4 | 3 | 1 | 0 |
| AI Not Currently Good Enough to Support Decisions | 19 | 2 | 10 | 7 | AI Indicates that There is No Right Answer | 4 | 0 | 3 | 1 |
| Not Useful Unless Integrates More Contextual Information | 17 | 9 | 6 | 2 | AI Helpful When Deciding Between Multiple Options | 4 | 0 | 2 | 2 |
| AI Potentially Useful for Less Experienced Providers | 15 | 8 | 3 | 4 | Desire for AI to Consider Other Disease Processes | 4 | 1 | 3 | 0 |
| AI Outputs Go Against Typical Practice | 14 | 8 | 5 | 1 | Desire to Know AI Accuracy | 4 | 2 | 1 | 1 |
| Desire for Literature References to Support AI Advice | 13 | 4 | 7 | 2 | AI Should Help Categorize Sepsis Patients | 4 | 3 | 1 | 0 |
| Desire for More Specific Advice | 12 | 3 | 5 | 4 | Concern Others May Over-Rely on AI Advice | 3 | 1 | 1 | 1 |
| Helpful to Reconsider Plan | 12 | 7 | 2 | 3 | AI Potentially Useful for Timing Interventions | 3 | 0 | 3 | 0 |
| Want AI to Double-Check Them | 11 | 4 | 6 | 1 | Potential for AI to Assess Volume Status | 3 | 0 | 3 | 0 |
| Options to Consider Potentially Useful | 10 | 2 | 7 | 1 | AI Unlikely to Help with Complex Cases | 3 | 0 | 1 | 2 |
| Continuous Evaluation is Important for Real Decisions | 10 | 5 | 4 | 1 | Options to Avoid Potentially Useful | 3 | 1 | 1 | 1 |
| Need to Know What Clinical Features the AI Uses | 10 | 7 | 2 | 1 | Concern AI Doesn't Capture Patient Nuances | 3 | 1 | 1 | 1 |
| Against AI Use for Sepsis Treatment | 8 | 3 | 4 | 1 | Desire to Control AI Inputs | 3 | 0 | 0 | 3 |
| Adds Burden | 8 | 0 | 3 | 5 | Not Helpful When Discordant | 3 | 1 | 2 | 0 |
| In Favor of AI Use for Sepsis Treatment | 8 | 4 | 1 | 3 | Desire for AI to Only Surface Evidence | 3 | 0 | 0 | 3 |
| Desire for AI to Descriptively Summarize Patient | 8 | 1 | 2 | 5 | Desire for Less Specific Advice | 3 | 0 | 0 | 3 |
| Helpful to Rethink Case | 7 | 5 | 2 | 0 | Helpful to Suggest New Alternatives | 3 | 2 | 1 | 0 |
| Simplicity of AI is Good | 7 | 3 | 2 | 2 | Desire for More Gradual Treatment Options | 3 | 1 | 1 | 1 |
| Helpful to Confirm Plan | 7 | 4 | 1 | 2 | Desire for More Nuanced Advice | 3 | 0 | 1 | 2 |
| AI Helpful for Complex Cases | 6 | 2 | 2 | 2 | Desire to Incorporate Other Goals Besides Survival | 3 | 1 | 2 | 0 |
| Not Useful Because Options Aren't Internally Consistent | 6 | 0 | 5 | 1 | Desire to See Options to Avoid After Making an Initial Decision | 2 | 0 | 0 | 2 |
| Not Useful Because It Suggests Plans They Are Not Considering | 5 | 0 | 1 | 4 | AI Potentially Useful for Managing Initial Resuscitation | 2 | 1 | 1 | 0 |
| Desire to Know What AI Risk is Relative To | 5 | 1 | 0 | 4 | Useful Regardless of Agreement | 2 | 0 | 2 | 0 |
| Desire for More Interactive Deliberative AI | 5 | 2 | 2 | 1 | AI Potentially Useful for Busy Services | 2 | 1 | 0 | 1 |
| Not Useful Because Treatment Options are Coupled | 5 | 0 | 1 | 4 | Desire to Know AI Confidence | 2 | 2 | 0 | 0 |
| Desire for More Precise Numeric Predictions | 5 | 0 | 4 | 1 | Concern for Lack of AI Accountability | 1 | 0 | 1 | 0 |
| AI Potentially Useful for Sepsis Screening | 5 | 0 | 4 | 1 | Skeptical of AI Assessing Volume Status | 1 | 1 | 0 | 0 |

## S.5 Patient Case Vignettes

The following sections provide the verbatim text of the vignettes provided to both the expert panel and the study participants. The numeric data was derived from the structured tables in MIMIC-IV, while the pseudonym and contextual details were written by the research team based on the patient's discharge note.

### S.5.1 Deborah Walker

Ms. Walker is a 64 year old woman with diabetes, hypertension, hyperlipidemia, and atrial fibrillation on warfarin is admitted to the ICU for treatment of sepsis and acute hypoxemic respiratory failure due to pneumonia. In the emergency department she was intubated and started on an IV norepinephrine drip.

You are evaluating her the morning after admission. Cultures of her blood, urine, and sputum are pending. She is receiving appropriate broad spectrum antibiotics. Her chest x-ray shows diffuse bilateral airspace opacities with a lower zone predominance. An echocardiogram shows a normal ejection fraction without significant valvular disease. On exam she is intubated and sedated with bilateral crackles in her lung fields.

The nurse gives you the following information:

- **HR:** 48 beats/minute (decreased from 60 four hours ago)
- **Heart rhythm:** atrial fibrillation
- **RR:** 15 breaths/minute (set rate = 14)
- **SaO2:** 98% on 40% FiO2 and 5 of PEEP
- **Norepinephrine dose:** 0.16 mcg/kg/min (decreased from 0.20 four hours ago)
- **Blood pressure:** 80/45 (decreased from 90/50 four hours ago)
- **MAP:** 57 (decreased from 63 four hours ago)
- **ABG:** 7.48/35/84
- **Serum lactate:** 1.9

- **Chemistries:**

| Lab | Value | Lab | Value |
|---|---|---|---|
| Sodium | 109 | Calcium | 7.1 |
| Potassium | 4.8 | Magnesium | 3.4 |
| Chloride | 76 | BUN | 15 |
| HCO3 | 24 | Creatinine | 0.8 |
| Glucose | 224 | | |

- **LFTs:** pending

In the last four hours she has received about 0.5 L of fluids. Her urine output is approximately 100 mL per hour, and she is net negative about 0.5 L since admission.

### S.5.2 Heather Thompson

Ms. Thompson is a 53 year old woman with Hepatitis C and cirrhosis is admitted to the ICU for treatment of sepsis with an unknown source. She resides in a skilled nursing facility and has an indwelling PICC line, placed a few weeks ago, to facilitate IV access. She has a history of bleeding esophageal varices and mild hepatic encephalopathy.

You are evaluating her the morning after admission. Cultures of her blood, urine, and ascitic fluid are pending. Her ascitic fluid was not suggestive of spontaneous bacterial peritonitis. She is receiving appropriate broad spectrum antibiotics. There is no clinical evidence of bleeding. Her chest x-ray shows mild interstitial edema but no infiltrate. A renal ultrasound was normal. On exam she is drowsy but arousable, and mildly dyspneic.

The nurse gives you the following information:

- **HR:** 56 beats/minute (58 four hours ago)
- **Heart rhythm:** sinus bradycardia

- **RR:** 17 breaths/minute (16 four hours ago)
- **SaO2:** 100% on 2L of oxygen by nasal cannula
- **BP:** 89/43 (decreased from 90/45 four hours ago)
- **MAP:** 58 (decreased from 60 four hours ago)
- **Chemistries:**

| **Lab** | **Value** | **Lab** | **Value** |
|---|---|---|---|
| Sodium | 128 | Calcium | 8.7 |
| Potassium | 5.1 | Magnesium | 2.8 |
| Chloride | 98 | BUN | 83 |
| HCO3 | 20 | Creatinine | 3.4 |
| Glucose | 70 | | |

- **Serum lactate:** 2.1 (first observation)

In the last four hours she has received 1.6 L of crystalloid. She is net positive 2.4 L since admission. She has made no urine since admission.

### S.5.3 Richard Sullivan

Mr. Sullivan is a 71 year old man with a history of heart failure with preserved ejection fraction, chronic renal insufficiency, coronary artery disease and diabetes is admitted to the ICU for treatment of sepsis due to acute cholecystitis. On admission he underwent percutaneous cholecystostomy with placement of a drainage tube. Following tube placement he experienced worsening respiratory failure requiring intubation, as well as worsening shock requiring intravenous norepinephrine and vasopressin.

You are evaluating him on day two of ICU admission. His blood cultures are growing gram negative rods, and he is being treated with appropriate antibiotics. There is no evident source of infection other than cholecystitis. He is sedated with propofol but follows commands when asked. On exam he has mild peripheral edema.

The nurse gives you the following information:

- **HR:** 76 beats/minute (increased from 67 four hours ago)
- **Heart rhythm:** sinus
- **RR:** 20 breaths/minute (set rate = 20)
- **SaO2:** 97% on 50% FiO2 and 15 of PEEP (increased from 10 since yesterday)
- **Plateau airway pressure on 6cc/kg/IBW =** 26 (decreased from 32 twelve hours ago)
- **Norepinephrine dose:** 0.04 mcg/kg/min (about the same since yesterday)
- **Vasopressin dose:** 0.04 units/min (on for the past day)
- **BP:** 100/52 (decreased from 115/61 twelve hours ago)
- **MAP:** 68 (decreased from 79 twelve hours ago)
- **ABG:** 7.43/40/103
- **Chemistries:**

| **Lab** | **Value** | **Lab** | **Value** |
|---|---|---|---|
| Sodium | 134 | Calcium | 8.1 |
| Potassium | 3.6 | Magnesium | 2.2 |
| Chloride | 101 | BUN | 78 |
| HCO3 | 26 | Creatinine | 1.5 |
| Glucose | 267 | | |

- **Serum lactate:** pending (3.9 on ICU admission)

Over the last four hours he has received 1.4 L of crystalloid. In the last 24 hours he has received 3 L of crystalloid. Concurrently, his PEEP has been increased from 10 to 15 in response to intermittent desaturations. His urine output is consistent at 40 ml/hr.

### S.5.4 Esther Goldberg

Ms. Goldberg is an 87 year old woman with a history of hypertension and hypercholesterolemia is admitted to the ICU for treatment of sepsis due to pneumonia complicated by new-onset atrial fibrillation. A CT angiogram of her chest confirmed her pneumonia and showed no evidence of pulmonary embolism. An EKG showed no evidence of ischemia and a cardiac troponin was normal. An echocardiogram showed normal systolic function.

You are evaluating her the morning after her admission. Blood cultures are pending, and she is being treated with appropriate antibiotics. Her chest x-ray shows diffuse bilateral infiltrates that have worsened since admission. She is receiving 100% oxygen by non-rebreather and is mildly dyspneic appearing. The nurse gives you the following information:

- **HR:** 107 beats/minute (decreased from 126 at ICU admission)
- **Heart rhythm:** atrial fibrillation
- **RR:** 20 breaths/minute (constant last eight hours)
- **SaO2:** 95% on 100% non-rebreather mask (decreased from 98% at ICU admission)
- **BP:** 84/45 (decreased from 101/53 four hours ago)
- **MAP:** 58 (decreased from 69 four hours ago)
- **Chemistries:**

| Lab | Value | Lab | Value |
|---|---|---|---|
| Sodium | 134 | Calcium | 7.5 |
| Potassium | 4.7 | Magnesium | 2.5 |
| Chloride | 101 | BUN | 11 |
| HCO3 | 30 | Creatinine | 0.6 |
| Glucose | 90 | | |

- **Serum lactate:** 0.8

Currently she is not receiving any intravenous fluids, and has been receiving a diuretic since admission. Her urine output is consistent 40 mL/hr, and she is net negative 1.2 L since admission.

### S.5.5 Margaret Kim

Ms. Kim is a 63 year old previously healthy woman who presents to the ICU for treatment of sepsis due to pyelonephritis. She presented to the emergency department with a two day history of fever, pyuria, and flank pain. A renal ultrasound showed no evidence of hydronephrosis, masses, calculi, or abscess.

You are evaluating her on arrival from the emergency department to the ICU. In the ED she received 1 L of IV crystalloid. Blood cultures are pending, and she is being treated with appropriate antibiotics. She is breathing 20 times a minute on room air with good oxygen saturation, but she appears mildly dyspneic and has faint crackles in her lower lung zones. The nurse gives you the following information:

- **HR:** 103 beats/minute
- **Heart rhythm:** sinus tachycardia
- **RR:** 20 breaths/minute
- **SaO2:** 96% on room air
- **BP:** 71/53
- **MAP:** 61
- **Chemistries:** pending
- **Serum creatinine:** pending
- **Serum lactate:** pending

**S.5.6 Mahesh “Mike” Patel**

Mr. Patel is a 51 year old male with a history of pancreatic cancer metastatic to the liver. He is admitted to the ICU for treatment of sepsis with an unknown source. He presented from home approximately a week after treatment with palliative chemotherapy. He is DNR/DNI, but he otherwise would like aggressive treatment so that he can return home and spend time with his family.

You are evaluating him the morning after ICU admission. A CT scan of his chest, abdomen, and pelvis shows no evidence of localizing infection. His physical exam is unremarkable. He has not made any urine since arrival. Blood cultures are pending and he is being treated with appropriate antibiotics. You are called to his bedside to see him for hypotension. The nurse gives you the following information:

- **HR:** 88 beats/minute (increased from 77 at admission)
- **Heart rhythm:** 1st degree AV block
- **RR:** 14 breaths/minute (decreased from 17 at admission)
- **SaO2:** 92% on room air (decreased from 97% at admission)
- **BP:** 75/45 (decreased from 79/52 at admission)
- **MAP:** 55 (decreased from 61 at admission)
- **Chemistries:**

| Lab | Value | Lab | Value |
|---|---|---|---|
| Sodium | 129 | Calcium | 6.6 |
| Potassium | 4.7 | Magnesium | 2.4 |
| Chloride | 94 | BUN | 81 |
| HCO3 | 20 | Creatinine | 4.5 |
| Glucose | 91 | | |

- **INR:** 2.9
- **Serum lactate:** pending

He has received 3 L of fluid overnight, but none in the last four hours. He had no urine output recorded overnight.

## References

1. Sivaraman V, Mason EP, Li ME, et al. Intelligent Reasoning Cues: A Framework and Case Study of the Roles of AI Information in Complex Decisions. arXiv:2602.00259 [cs]. 2026. doi: 10.1145/3772318. 3790953. url: http://arxiv.org/abs/2602.00259 (visited on 02/07/2026).
2. Johnson A, Bulgarelli L, Pollard T, Horng S, Celi LA, and Mark R. MIMIC-IV (version 1.0). 2020.
3. Komorowski M, Celi LA, Badawi O, Gordon AC, and Faisal AA. The Artificial Intelligence Clinician learns optimal treatment strategies for sepsis in intensive care. Nature Medicine 2018;24:1716–20.
4. Nanayakkara T, Clermont G, Langmead CJ, and Swigon D. Unifying cardiovascular modelling with deep reinforcement learning for uncertainty aware control of sepsis treatment. PLOS Digital Health 2022;1:e0000012.
5. Sivaraman V, Bukowski LA, Levin J, Kahn JM, and Perer A. Ignore, Trust, or Negotiate: Understanding Clinician Acceptance of AI-Based Treatment Recommendations in Health Care. CHI ’23 2023.